\documentclass[fleqn,usenatbib]{mnras}

\usepackage[T1]{fontenc}
\usepackage{rotating}
\usepackage{autobreak}
\usepackage{hyperref}
\usepackage{caption}

\allowdisplaybreaks
\DeclareRobustCommand{\VAN}[3]{#2}
\let\VANthebibliography\thebibliography
\def\thebibliography{\DeclareRobustCommand{\VAN}[3]{##3}\VANthebibliography}

\usepackage{graphicx}	
\usepackage{amsmath}	
\usepackage{amssymb}	

\title[X-ray reprocessing in GX 301-2]{X-ray reprocessing in accreting pulsar GX 301-2 observed with \textit{Insight}-HXMT}

\author[Long, Ji et al.]{%
	L. Ji$^{1}$ \thanks{E-mail: ji.long@astro.uni-tuebingen.de},
    V. Doroshenko$^{1, 6}$,
    V. Suleimanov$^{1, 4, 6}$,
	A. Santangelo$^{1, 3}$,
	M. Orlandini$^{9}$ 
    \newauthor
    J., Liu$^{10}$ 
    L. Ducci$^{1, 2}$,
	S. N. Zhang$^{3,7}$,
	A. Nabizadeh$^{5}$,
	D. Gavran$^{1}$
% ...
\\
$^{1}$ Institut f\"ur Astronomie und Astrophysik, Kepler Center for Astro and Particle Physics, Eberhard Karls Universit\"at, Sand 1,\\ 72076 T\"ubingen, Germany\\
$^{2}$ ISDC Data Center for Astrophysics, Universit\'e de Gen\`eve, 16 chemin d'\'Ecogia, 1290 Versoix, Switzerland\\
$^{3}$ Key Laboratory for Particle Astrophysics, Institute of High Energy Physics, Beijing 100049, China\\
$^{4}$ Kazan (Volga region) Federal University, Kremlevskaya str. 18, 42008 Kazan, Russia,\\
$^{5}$ Department of Physics and Astronomy, University of Turku, FI-20014 Turku, Finland\\
$^{6}$ Space Research Institute of the Russian Academy of Sciences, Profsoyuznaya Str. 84/32, Moscow 117997, Russia\\
$^{7}$ University of Chinese Academy of Sciences, Chinese Academy of Sciences, Beijing 100049, People's Republic of China\\
$^{8}$ Department of Astronomy, Beijing Normal University, Beijing 100088, People's Republic of China\\
$^{9}$ INAF – Osservatorio di Astrofisica e Scienza dello Spazio di Bologna, Via Piero Gobetti 101, I-40129 Bologna, Italy\\
$^{10}$ Beijing Planetarium, 138 Xizhimenwai Road, Beijing 100044, China\\
$^{11}$ Department of Astronomy, Tsinghua University, Beijing 100084, China\\
$^{12}$College of physics Sciences \& Technology, Hebei University, No. 180 Wusi Dong Road, Lian Chi District, Baoding City,\\
Hebei Province, 071002 China\\
$^{13}$School of Physics and Optoelectronics, Xiangtan University, Yuhu District, Xiangtan, Hunan, 411105, China\\
}

\date{Accepted XXX. Received YYY; in original form ZZZ}

\pubyear{2020}

\begin{document}
\label{firstpage}
\pagerange{\pageref{firstpage}--\pageref{lastpage}}
\maketitle
\begin{abstract}
We investigate the absorption and emission features in observations of GX 301-2 detected with \textit{Insight}-HXMT/LE in 2017-2019. 
At different orbital phases, we found prominent Fe K$\alpha$, K$\beta$ and Ni K$\alpha$ lines, as well as Compton shoulders and Fe K-shell absorption edges.
These features are due to the X-ray reprocessing caused by the interaction between the radiation from the source and surrounding accretion material.
According to the ratio of iron lines (K$\alpha$ and K$\beta$), we infer the accretion material is in a low ionisation state.
We find an orbital-dependent local absorption column density, which has a large value and strong variability around the periastron. 
We explain its variability as a result of inhomogeneities of the accretion environment and/or instabilities of accretion processes.
In addition, the variable local column density is correlated with the equivalent width of the iron K$\alpha$ lines throughout the orbit, which suggests that the accretion material near the neutron star is spherically distributed.
\end{abstract}




\section{Introduction}
In high mass X-ray binaries (HMXBs), the main component of the mass outflow emitted by the donor star and responsible for the X-ray emission in the vicinity of the accreting compact object can be a spherically symmetric wind, a circumstellar disk, or a gas stream.
The radiation, as seen from Earth, is absorbed both in the interstellar medium (ISM) and within the binary system.
The latter is attributed to the accretion material that surrounds the compact star which is often inhomogeneous and highly clumpy as reflected by highly variable absorption.
Absorption and re-emission of X-rays are affected by the distribution of material and one of the key diagnostic tools to probe the environment in binary systems \citep[e.g.,][]{Aftab2019}.
Fluorescence lines, especially of iron atoms, are prominent features of the X-ray reprocessing in HMXBs \citep[see, e.g.,][]{Torrejon2010, Tzanavaris2018}. 
They are produced by the absorption of high energy photons that remove K-shell electrons and lead to electronic transitions (L$\rightarrow$K: Fe $\rm K\alpha$ and M$\rightarrow$K:  Fe $\rm K\beta$) \citep{Kallman2004}.
In addition, when the compact star is embedded in a dense wind, the Compton scattering is non-negligible, which scatters a fraction of emissions out of the line-of-sight (LOS) and reduces the observed flux.
On the other hand, the down-scattering of fluorescence lines may lead to the appearance of a 'Compton shoulder' (CS) due to electron recoils \citep{Matt2002,Watanabe2003}.

GX 301-2 is an HMXB  consisting of a highly magnetized ($B\sim4\times10^{12}$\,G, or even larger \citealt{Doroshenko2010}) pulsar and a B-type hyper-giant star Wray 977 \citep{Vidal1973, Kaper1995,Staubert2019}.
According to modelling of high-resolution optical spectra, Wray 977 has a mass of 43$\pm$10 $M_{\odot}$, a radius of 62 $R_{\odot}$ and looses mass through powerfull stellar winds at a rate of $\sim10^{-5}M_\odot\,{\rm yr}^{-1}$ with terminal velocity of 300\,$\rm km\,s^{-1}$ \citep{Kaper2006}.
The system is highly eccentric ($e\sim$ 0.46), with an orbital period of $\sim41.5$\,d, and exhibits strong variation of the X-ray flux with orbital phase \citep{Koh1997,Doroshenko2010}.
In particular, periodic outbursts at the orbital phase $\sim$1.4\,days before the periastron passage \citep{Sato1986}, and a fainter one near the apastron passage are observed \citep{Pravdo1995}. 
The broad-band X-ray spectrum is orbital phase-dependent and can be approximately described as a power-law with a high energy cutoff and a cyclotron resonant scattering feature (CRSF) around 40\,keV \citep{Kreykenbohm2004,Mukherjee2004,LaBarbera2005,Doroshenko2010,Suchy2012,Islam2014,Furst2018,Nabizadeh2019}.
During the periastron flares the source exhibits strong variability with an amplitude of up to a factor of 25, reaching a few hundreds mCrab in the energy band of 2-10\,keV \citep[e.g.,][]{Rothschild1987, Pravdo1995}.
The flares are accompanied by the variability of the equivalent hydrogen column density ($\rm N_{\rm H}$) and of the fluorescent iron lines, which is believed to be associated with clumpiness of the stellar wind, launched from the donor star \citep{Mukherjee2004}.
We note the \textit{clumpiness} in this paper refers to any inhomogeneities in the stellar wind/stream, which are higher density regions, regardless of its specific formation mechanisms.
On the other hand, \citet{Furst2011} reported a long \textit{XMM-Newton} observation in GX 301-2 around its periastron, which also exhibits systematic variations of the flux and $\rm N_{H}$ at a time-scale of a few kilo-seconds.
Several wind accretion models, consisting of stellar winds and a gas stream, were proposed to explain the observed flares \citep[e.g.,][]{Haberl1991,Leahy1991, Leahy2008, Monkkonen2020}.

As already mentioned, reprocessing of X-ray emission can be used to probe the environment surrounding the neutron star.
We note that, however, a comprehensive and detailed study of the X-ray reprocessing over the entire orbit is still missing.
Thanks to the high cadence observations of \textit{Insight-Hard X-ray Modulation Telescope (HXMT)} \citep{Zhang2019} in 2017-2019, we are able to study the X-ray reprocessing at different orbital phases, and compare it with the result of the flaring episode.
This work aims at improving our understanding on the accretion environment of GX 301-2 by studying X-ray processing.
This paper is organised as follows: In Section 2, we describe observations and procedures adopted for data reduction; the spectral analysis and results are represented in Section 3; we summarise our conclusions in Section 4 and 5.
\begin{figure}
	\centering
	\includegraphics[width=3.2in]{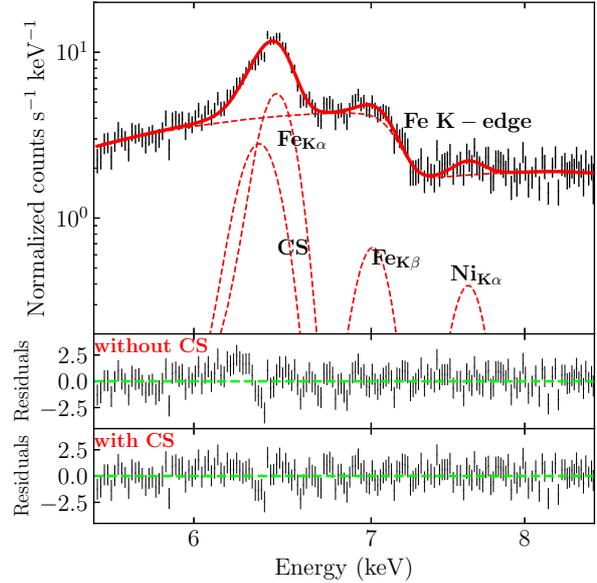}
	\caption{A representative spectrum of GX 301-2 in our sample (obsID:P010130901601; upper panel). The spectrum is fitted in the energy range of 5.5-8.5\,keV with an absorbed (tbabs) powerlaw model including three emission lines, i.e., Fe K$\alpha$, Fe K$\beta$ and Ni K$\alpha$, and an additional Compton shoulder described as a box function.
	We show residuals with and without the CS component in the two bottom panels.	
}
	\label{fig:example}
\end{figure}

\begin{figure}
	\centering
	\includegraphics[width=3.2in]{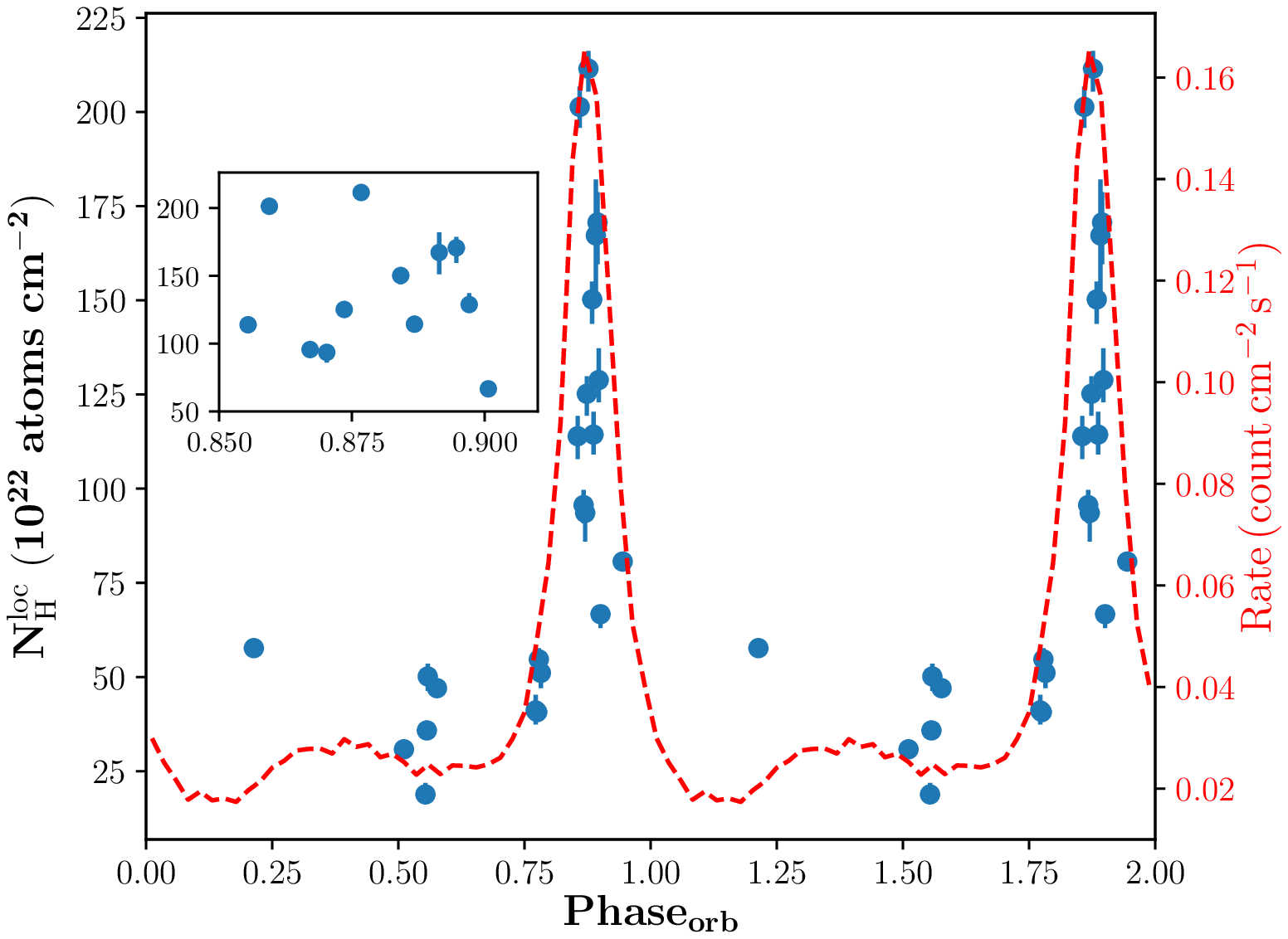}
	\includegraphics[width=3.2in]{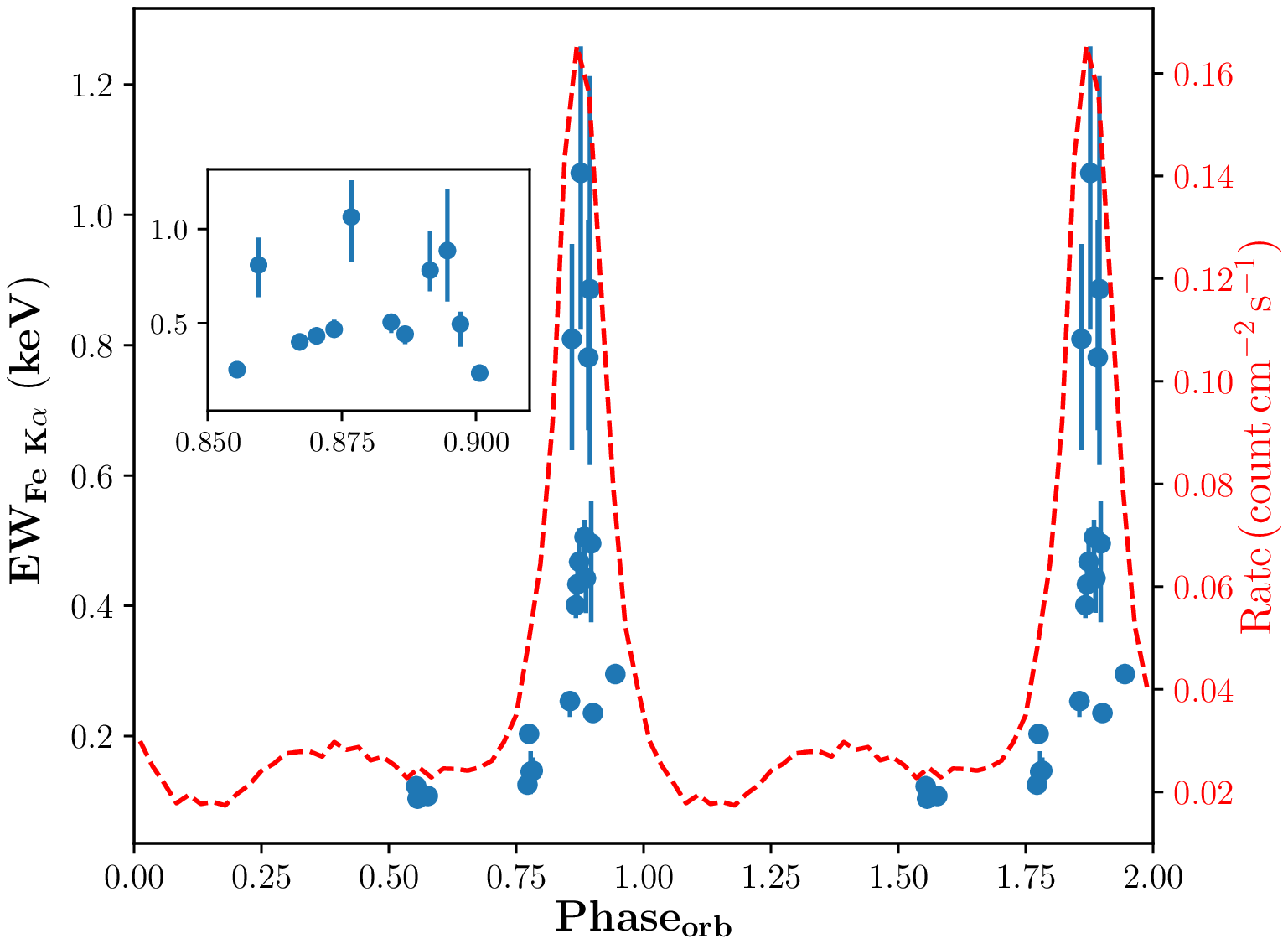}
	\includegraphics[width=3.2in]{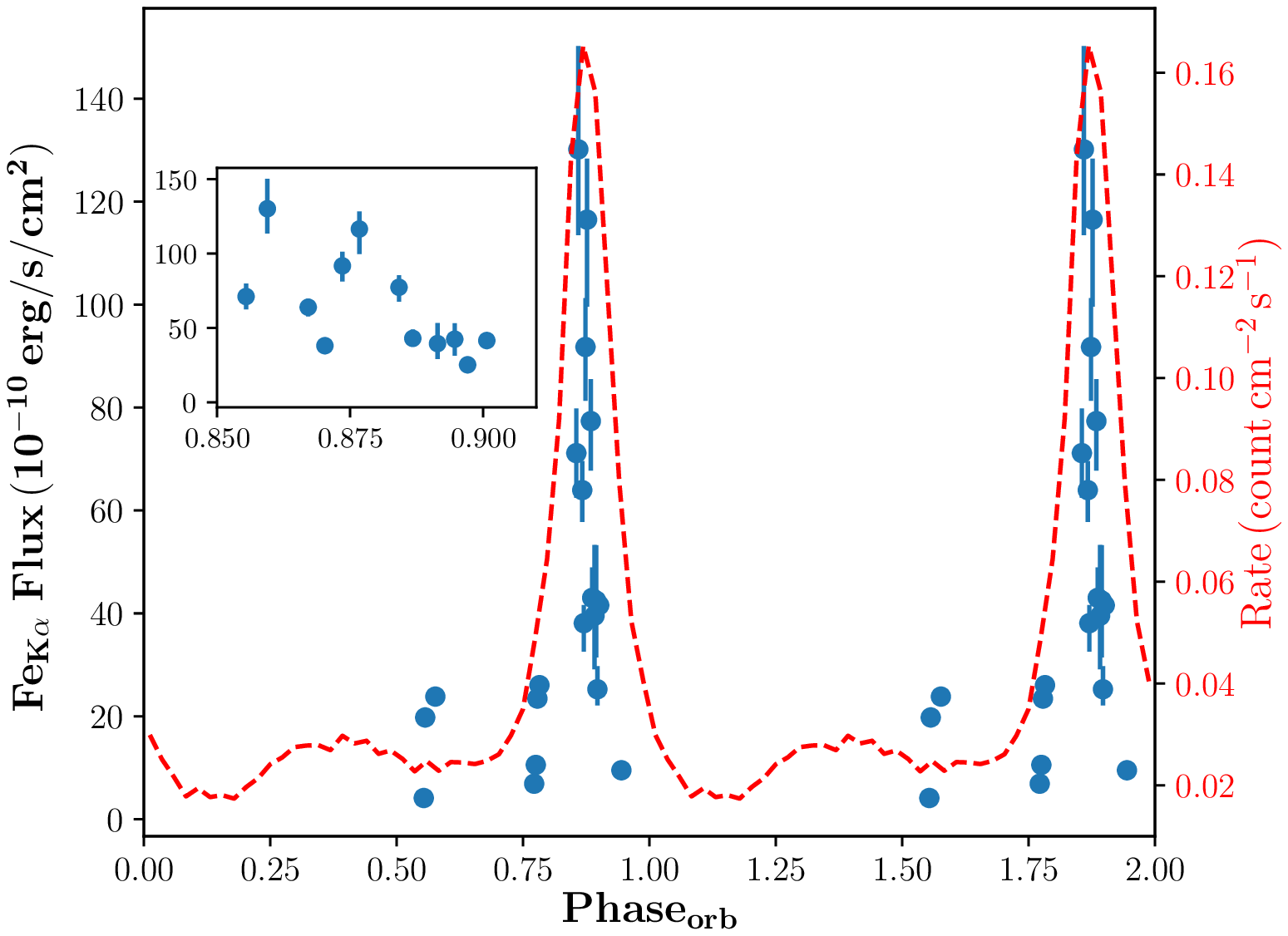}
	\caption{The equivalent local hydrogen column ($N_{\rm H}^{\rm loc}$) vs. the  orbital phase (upper panel). 
	 The inset is a zoom-in to show the significant $N_{\rm H}^{\rm loc}$ variation around the orbital phase 0.9.
	 The orbital modulation of the flux (red) is also superimposed, which is monitored by \textit{Swift}/BAT in the energy range of 15-50\,keV. Here the binary ephemeris is adopted from \citet{Koh1997}.
	 We show orbital evolutions of EWs and intensities of Fe K$\alpha$ lines in the middle and bottom panels.    
 }
	\label{fig:nH}
\end{figure}

\begin{figure}
	\centering
	\includegraphics[width=3.2in]{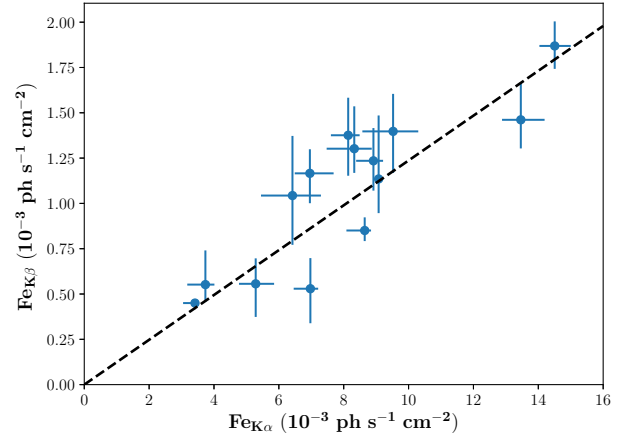}
	\caption{Strengths of Fe K${\alpha}$ and Fe K${\beta}$ lines, and their linear relation (the black dashed line).}
	\label{fig:FealphaBeta}
\end{figure}

\begin{figure}
	\centering
	\includegraphics[width=3.2in]{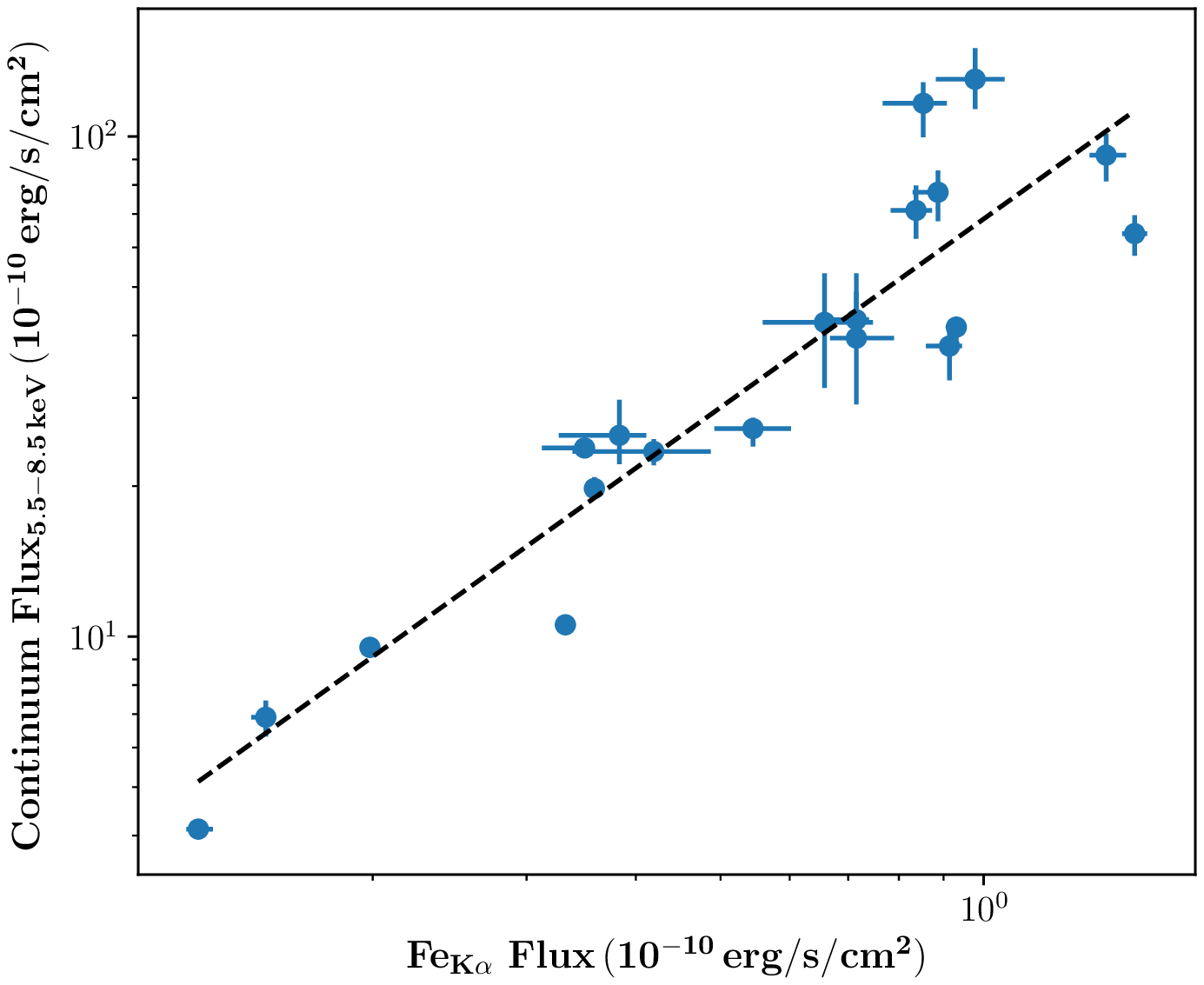}
	\includegraphics[width=3.2in]{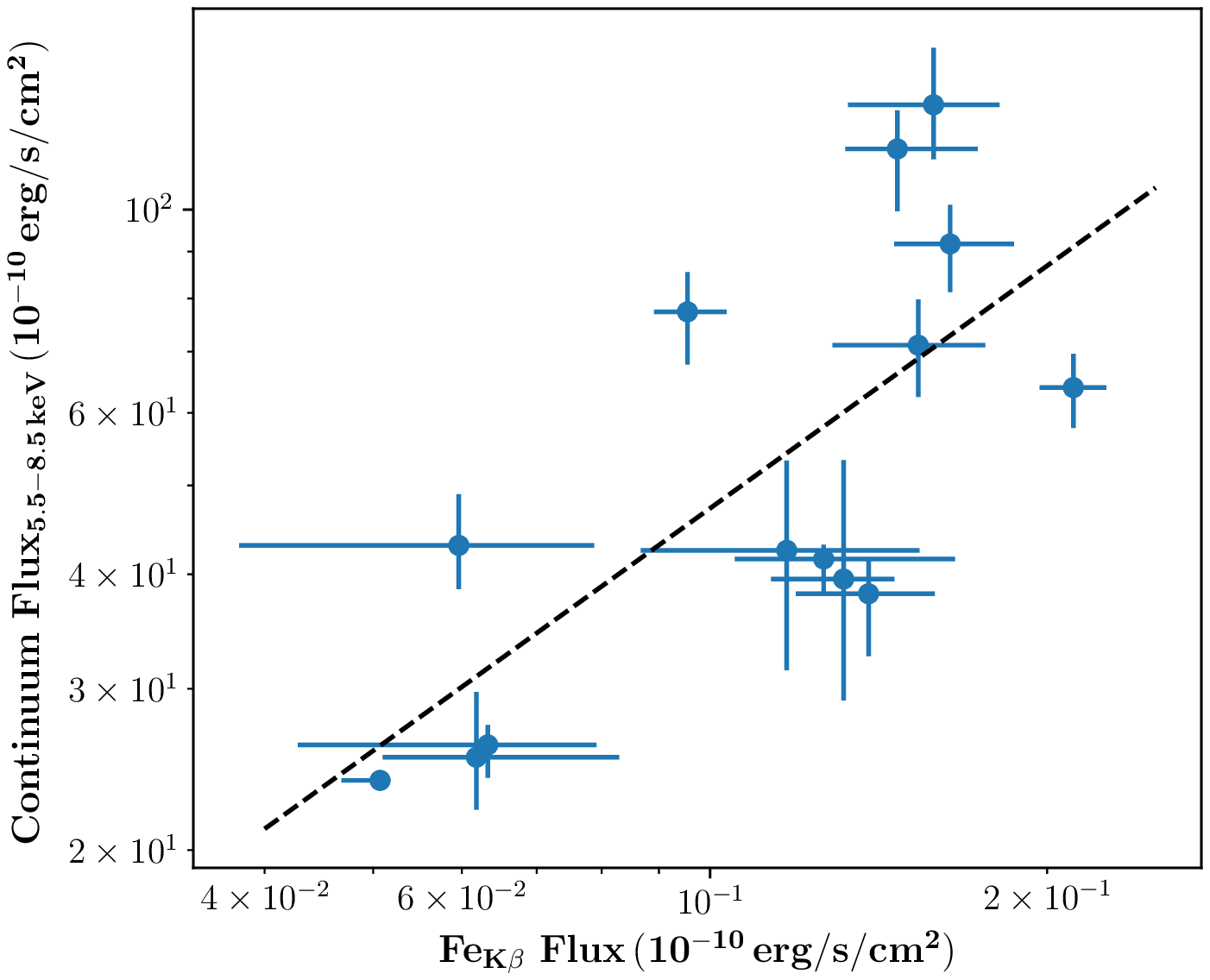}
	\caption{The upper (lower) panel shows the relation between the $\rm Fe\ {K\alpha}$ ($\rm Fe\ {K\beta}$) flux and the unabsorbed flux of the continuum in the energy range of 5.5-8.5\,keV. Dashed lines represent linear fits in the logarithmic space, which result in linear coefficients of 1.25 and 0.88 for $\rm Fe\ {K\alpha}$ and $\rm Fe\ {K\beta}$ lines, respectively.	
	}
	\label{fig:Fe_vs_con}
\end{figure}

\begin{figure}
	\centering
	\includegraphics[width=3.2in]{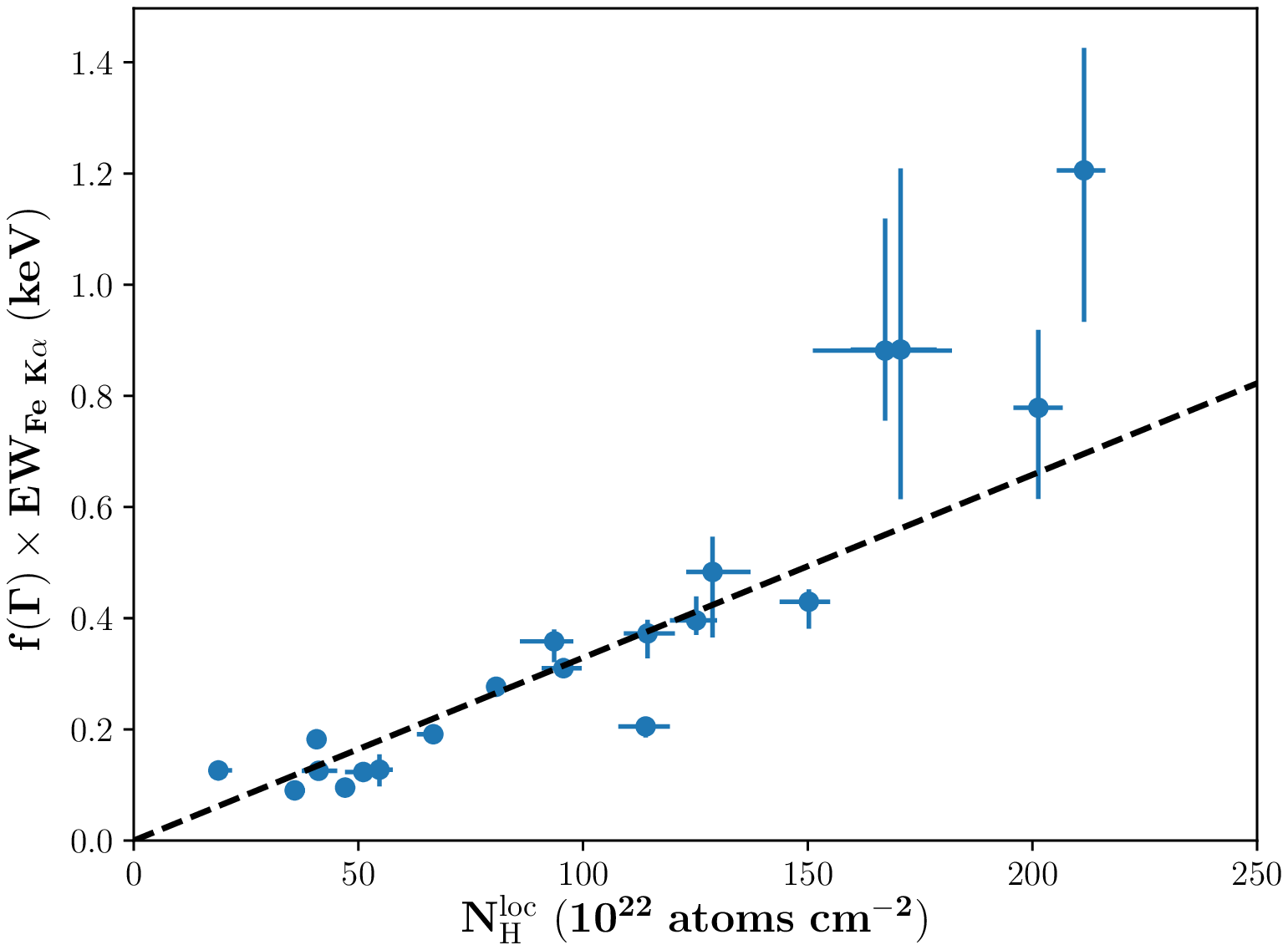}
	\includegraphics[width=3.5in]{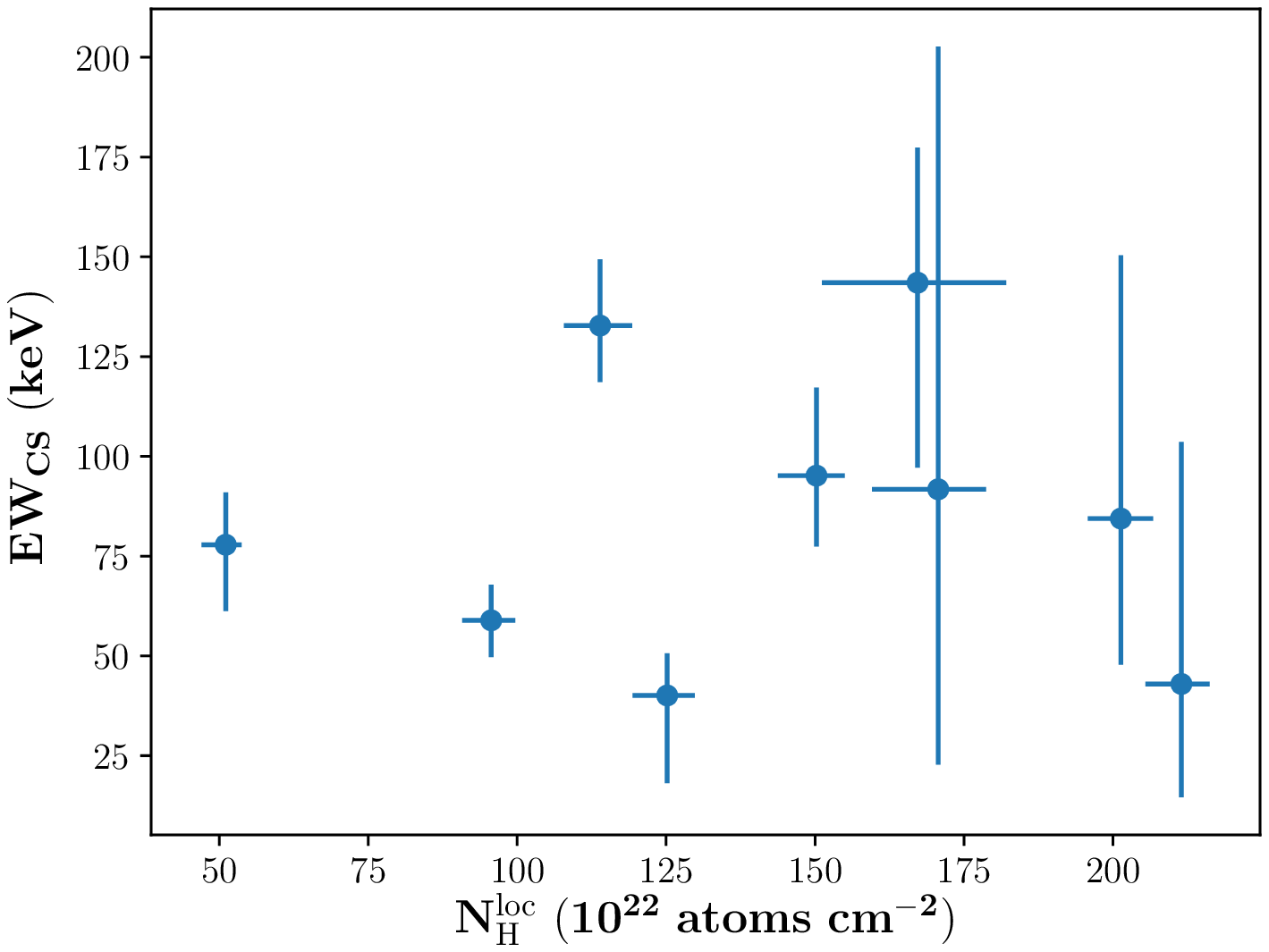}
	\caption{The upper (bottom) panel shows equivalent widths of Fe K$\alpha$ lines (CSs) as a function of the local absorption column $N_{\rm H}^{\rm loc}$. The black dashed line represents the theoretical estimation ($\rm EW_{Fe\ K\alpha}(eV) \approx 3\times10^{-22} N_{H}$) assuming a spherical shell of gas surrounding a point source of continuum radiation \citep{Kallman2004}.
	}
	\label{fig:nHCS}
\end{figure}
\section{Data reduction}
\textit{Insight}-HXMT is the first Chinese X-ray satellite, which consists of three telescopes, i.e., the low, medium and high energy telescopes \citep{Zhang2014,Zhang2019}.
In this work, we only used the low energy telescope (LE), which is made up of swept charge devices (SCDs) and cover the energy range of 1-10\,keV.
The broad-band spectral analysis will be published elsewhere, and here we only focus on the narrow emission and absorption features (i.e., iron line complex) observed in the soft X-ray band, and associated them with reprocessing of X-ray emission by winds. 
LE has an energy resolution of 140\,eV at 5.9 keV, and an effective area of 384\,$\rm cm^2$ \citep{Chen2019}.
Thanks to its fast-readout, LE does not suffer from photon saturation and pile-up effects, and is therefore capable of observing strong sources, like the flaring state of GX~301-2.
\textit{Insight}-HXMT performed 67 pointing observations between 2017-2019, with an averaged exposure of $\sim$ 2000\,s. 
However, in some cases, the observations were dominated by the background, especially when the source was faint, which precludes detailed spectral analysis. In particular, we only select the observations in which the source contributes to more than 25\% of the count rate between 1-10\,keV, and the exposure is larger than 1000\,s for the analysis.
Here the background was estimated by using {\sc lebkgmap}, a {\sc python} code that has been included in HXMTDAS-2.02.1.
As a result, 23 observations have been selected in our sample, and their summary is shown in Table~\ref{tab:ObsID}.
We performed the data reduction of \textit{Insight}-HXMT, following the official user guide \footnote{see \url{http://enghxmt.ihep.ac.cn/SoftDoc/169.jhtml}}: the criteria for data screening is that the elevation angle > 10 degree; the geomagnetic cutoff rigidity > 8\,GeV; the pointing offset angle < 0.1 degree; at least 300\,s away from the South Atlantic Anomaly (SAA).
During the spectral analysis, we used {\sc xspec v12.10}, an X-ray spectral fitting package in {\sc heasoft v6.26} \citep{Arnaud1996}.
We assumed solar abundances \citep{Wilms2000}, as suggested by previous observations \citep{Torrejon2010}. 
In this paper, we identified the existence of a component only if its detection is at a confidence level of > 3\,$\sigma$.
The significance was estimated by Monte-Carlo simulations performed by using {\sc simftest}, a built-in script in {\sc xspec}.
A systematic error of the LE calibration (CALDB 2.02.01), regarding the Energy-to-Channel (E-C) relation, has been found recently \footnote{private communication with Dr. Xiaobo Li}, which will lead to an overestimation of the line energy by a few tens of eV.
Such a deviation is expected to be stable overall observations.
Therefore an offset of the E-C relation has been taken into account in our analysis (see below).
All uncertainties in this paper correspond to a 68 \% confidence level.

\begin{figure}
	\centering
	\includegraphics[width=3.2in]{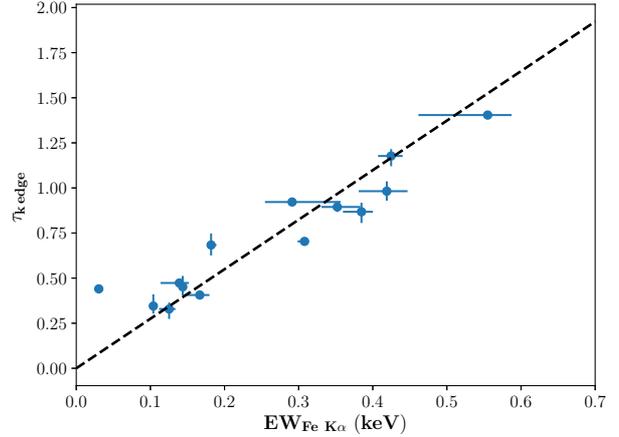}
	\caption{The relationship between the equivalent width of the iron $\rm K_{\rm \alpha}$ line and the optical depth of the iron K-edge absorption, where the black line represents a linear fitting.}
	\label{fig:tau}
\end{figure}

\section{Results}
We focused on the energy band of 5.5-8.5\,keV, where the emission and absorption features are expected from past observations of the source with \textit{Chandra}/HEG  \citep{Watanabe2003,Torrejon2010, Tzanavaris2018}. 
In this narrow energy band, the continuum can be described with a power-law spectral model.
The model ($M$) we used is:
\begin{equation}
\begin{aligned}
M = &{\rm abs1(N_{\rm H}^{\rm gal}) \times \big \{ abs2(N_{\rm H}^{\rm loc}) \times cabs(N_{\rm H}^{\rm loc}) } \\
& {\rm  \times powerlaw + Gaussian\ lines + box \big \} }\\               
\end{aligned}
\label{equ1}
\end{equation}
where
\begin{equation}
{\rm box} = \left\{ \begin{array}{l}
{\rm norm, when\ 6.24\,keV<E<6.40\,keV}\\
{\rm 0, otherwise}
\end{array} \right.
\label{box}
\end{equation}
, and $\rm abs1(N_{\rm H}^{\rm gal})$ and $\rm abs2(N_{\rm H}^{\rm loc})$ represent photoelectric absorption caused by the Galactic and local interstellar medium respectively.
In practice, we adopt the \texttt{tbnew} \footnote{\url{https://pulsar.sternwarte.uni-erlangen.de/wilms/research/tbabs/}}
 model to describe them. 
The $\rm N_{\rm H}^{\rm gal}$ was fixed at $\rm 1.4 \times 10^{22}\,atoms\ cm^{-2}$ \citep{HI4PI2016}.
The \texttt{cabs}\footnote{\url{https://heasarc.gsfc.nasa.gov/xanadu/xspec/manual/node234.html}} model was used to account for the Compton scattering that decreases the flux along the line of sight.
We note that our data can be also well described without inclusion of the \texttt{cabs} component which is coupled with the normalization of the powerlaw component and does not affect any of our conclusions. 
We also included several Gaussian lines to model fluorescent lines, and the line widths were fixed at 1\,eV
\footnote{
	We confirmed that LE is not able to constrain the line widths and only upper limits can be obtained if setting up them free.}
, as suggested by \textit{Chandra}/HEG.
Only most prominent fluorescent lines, i.e., Fe K$\alpha$, Fe K$\beta$ and Ni K$\alpha$, were considered because of statistics.
In addition, we found that sometimes the Fe K$\alpha$ line deviates from the symmetric Gaussian line profile (see residuals in Figure~\ref{fig:example}), which is likely due to the presence of a Compton shoulder.
To account for this feature we included also a box-shaped function covering the energy range 6.24-6.40\,keV \citep[for details, see][]{Matt2002}.
We considered the presence of CSs only if the box function could improve the goodness-of-fit at a significance level of > 3$\sigma$ estimated by {\sc simftest}.
The model in Eq.~\ref{equ1} can well describe all spectra in our sample with an averaged reduced-$\chi^2$ of 0.9 (269 dof), and we show a representative example in Figure~\ref{fig:example}.
As studied by \citet{Torrejon2010}, recombination lines (\ion{Fe}{XXV} and \ion{Fe}{XXVI}) were also detected in some high mass X-ray binaries. However, in our sample we did not find signals of \ion{Fe}{XXV} and \ion{Fe}{XXVI}, which is consistent with the previous conclusion that recombination lines are not present in supergiant X-ray binaries \citep{Gimenez2015}.

It has been known that the $N_{\rm H}^{\rm loc}$ in GX 301-2 exhibits short-term and long-term variability \citep{Mukherjee2004, Furst2011}.
The former is thought to be caused by the clumpiness of the accretion material, while the latter is attributed to the evolution of the accretion environment of stellar winds or the gas stream, as a function of the orbital phase.
\textit{Insight}-HXMT covered a substantial fraction of the orbit, which allows us investigating the $N_{\rm H}^{\rm loc}$ variability at different timescales. The results are presented in Figure~\ref{fig:nH} and Table~\ref{tab:par1}. As it is evident from the figure, $N_{\rm H}^{\rm loc}$ is largest and exhibits strongest variability around the orbital phase 0.9, when the flux is enhanced as well although it is important to emphasise that the local absorption column remains variable throughout the orbit.
In addition, observational IDs having a same prefix (such as P0101309001*) are parts of a long pointed observation separated by 
short gaps, which allows us studying the $N_{\rm H}^{\rm loc}$ variation at a short time-scale of a few kilo-seconds.
To conclude, the observed $N_{\rm H}^{\rm loc}$ is significantly variable at this short time-scale, which indicates that the $N_{\rm H}^{\rm loc}$ fluctuation in Figure~\ref{fig:nH} is mainly caused by short-term effects. 
This result is well consistent with the long exposure \textit{XMM}-Newton observation \citep{Furst2011}.

The centroid energy of the Fe K$\alpha$ line ($\sim$ 6.43\,keV) appears to be consistent for all observations and corresponds to the ionization degree of more than $\ion{Fe}{XVIII}$ \citep{Kallman2004}.
However, we note that the ionization state might be overestimated considering the optical depth of the absorber, and is inconsistent with previous reports
\citep[e.g.,][]{Watanabe2003, Furst2011}.
On the other hand, if materials are ionized beyond $\ion{Fe}{XVIII}$, the resulting photoionization threshold should be much higher \citep[Fig. 4c in][]{Kallman2004}, which is inconsistent with the fact that the data can be modelled with an absorption edge when assuming neutral ions (see below).
We note that this discrepancy is likely caused by the systematic uncertainty of the E-C relation as mentioned above.
Therefore, in practice, we fitted the spectrum of the first observation (ObsID:P010130900101), by freezing Fe K$\alpha$ line at 6.4\,keV and setting up the {\it gain offset}\footnote{For details, see \url{https://heasarc.gsfc.nasa.gov/xanadu/xspec/manual/node99.html}.} free.
The resulting offset was 35\,eV, which hereafter was considered for other spectra.

The fitting results have been summarized and tabulated in Table~\ref{tab:par1}.
Similar to the $N_{\rm H}^{\rm loc}$, the Fe lines are also highly variable and related to the orbital phase (Figure~\ref{fig:nH}). 
Both the equivalent width (EW) and the flux of Fe K$\alpha$ lines present large values and strong variability around the periastron.
We show the linear relation between intensities of Fe K$\beta$ and K$\alpha$ lines in Figure~\ref{fig:FealphaBeta}.
A linear fitting leads to a linear coefficient of 0.12$\pm$0.01.
We note that, in theory, this coefficient is described as a function of the iron ionization state \citep[see Fig. 2 in][]{Palmeri2003}. 
Therefore, this result suggests a low ionization state (electron occupancy $\gtrsim$ 22) and is consistent with our assumption mentioned above.
In Figure~\ref{fig:Fe_vs_con} we represent the unabsorbed flux of the continuum in the energy range of 5.5-8.5\,keV against the flux of Fe K$\alpha$/K$\beta$ lines.
Clearly both of them are positively correlated, which is consistent with the relation among other sources  \citep[see Fig.~6 in][]{Gimenez2015}.

Assuming that Fe K$\alpha$ lines are emitted from a spherical shell of gas surrounding the source, the equivalent width is expected to be related to the local hydrogen column density as $\rm EW_{Fe\ K\alpha}(eV) \approx 3\times10^{-22} N_{\rm H}^{\rm loc}$, if considering a power-law continuum spectrum with a photon index ($\Gamma$) of 2 \citep{Kallman2004,Inoue1985}.
Therefore, we compared our observations with the theoretical prediction shown in the upper panel of Figure~\ref{fig:nHCS}.
Since the photon index does not strictly equal to 2 in real observations, we included a correction factor f($\Gamma$), i.e., a normalization factor with respect to the case of $\Gamma$=2 \citep{Endo2002}, where
\begin{equation}
{\rm{f(}}\Gamma {\rm{) = }}\frac{{\int_{{{\rm{E}}_{{\rm{th}}}}}^\infty  {{{\rm{E}}^{ - 2}}{\sigma_{{\rm{Fe}}}}{\rm{(E)dE}}} }}{{\int_{{{\rm{E}}_{{\rm{th}}}}}^\infty  {{{\rm{E}}^{ - \Gamma }}{\sigma_{{\rm{Fe}}}}{\rm{(E)dE}}} }}
\end{equation}
Here $\rm{E}_{\rm{th}}$ is the photoionization threshold energy, and $\sigma_{{\rm{Fe}}}$ is the K shell photoionization cross-section.
In spite of some scattering points, the EW significantly increases with the increasing of $N_{\rm H}^{\rm loc}$, which is generally in agreement with the theoretical estimation.
The slight departure from the expected linear relationship, e.g., for the points with large $N_{\rm H}^{\rm loc}$ values, might be a hint of the aspherical accretion, such as the formation of the accretion disk suggested by \citet{Monkkonen2020}.

In nine out of 23 observations the Compton shoulder (CS) appears to be statistically significant. We show the relation between the EW of this component and the corresponding $N_{\rm H}^{\rm loc}$ in the bottom panel of Figure~\ref{fig:nHCS}. 
No apparent correlation can be identified due to the large scatter of individual data points. Considering the strong dependency between CSs and strong Fe K$\alpha$ lines, we caution that such a scattering might be due to the imperfect description of observed features with simplified spectral models.
Further observations with a better energy resolution are strongly encouraged in the future. 

As shown in Figure~\ref{fig:example}, a flux decrease is visible around 7\,keV, which is attributed to the K-shell photoionization.
It is useful to determine the relation between $N_{\rm H}^{\rm loc}$ and the absorption optical depth.
We performed an additional spectral analysis using the same model mentioned above but setting up the iron abundance of "abs2" at 0.
In addition, we included a phenomenological model, i.e., "edge"\footnote{\url{https://heasarc.gsfc.nasa.gov/xanadu/xspec/manual/node236.html}} in {\sc xspec}, to describe the absorption structure.
There are two parameters in the "edge" model, the threshold energy ($E_{\rm c}$) and the absorption optical depth ($\tau_{\rm K\,edge}$).
To avoid the coupling with Fe K$\beta$ lines that are around 7.1\,keV, we fixed $E_{\rm c}$ at 7.11\,keV \footnote{\url{http://skuld.bmsc.washington.edu/scatter/AS\_periodic.html}}, i.e., the predicted value for neutral irons.
We show the relation and a linear fitting, i.e., $\tau_{\rm K\,edge} = (2.5\pm0.4)\,{\rm \frac{EW_{\rm Fe\,K\alpha}}{\rm 1\,keV} }$, in Figure~\ref{fig:tau}.
This relation suggests that the reprocessing material reaches an optical depth unit for EW$\sim$ 400\,eV.

\section{Discussion}
As illustrated in Fig.~\ref{fig:nH}, a strong variability of the local absorption column at all orbital phases is revealed by \textit{Insight}-HXMT observations.
In general, the $N_{\rm H}^{\rm loc}$ shows a considerable increase close to the periastron ($\phi$=0.9) of the binary system, which is consistent with previous reports \citep[e.g.,][]{Haberl1991, Mukherjee2004}.
Several models have been proposed to explain the observed dependence of both the absorbtion and flux on the orbital phase \citep{Stevens1988, Haberl1991, Leahy1991, Leahy2008, Monkkonen2020}.
It is generally believed that the pulsar accretes matters both via a spherically symmetric wind \citep{Castor1975} and focused accretion stream with a higher density.
The latter is believed to be due to the enhanced mass loss on the surface of the supergiant star towards the neutron star, resulting in an Archimedes spiral-like structure because of the conservation of angular momentum \citep{Leahy2008}.
We note that, although such a hybrid model can successfully reproduce the flux modulation of this source, the expected maximum of $N_{\rm H}^{\rm loc}$ is at the orbital phase $\phi$=0.2, which is inconsistent both with the \textit{RXTE} results and our observations \citep[Fig. 6 in][]{Leahy2008}. On the other hand, the presence of the accretion stream itself with a shape similar to that deduced by \cite{Leahy2008} has been directly confirmed through near infrared interferometry \citep[see i.e. Fig. 13 in ][]{2017ApJ...844...72W}, which might suggest that the observed absorption is not strongly affected by global distribution of matter within the binary system and is mostly local. This conclusion is also supported by the fact that the local absorption column observed by \textit{Insight}-HXMT is strongly variable on short timescales at all orbital phases, which is also consistent with reports from the literature \citep{Mukherjee2004,Furst2011}.

In general, two possible interpretations have been discussed in the literature to account for the observed variability of $N_{\rm H}^{\rm loc}$ in high mass X-ray binaries: wind clumps due to the line-deshadowing instability \citep[e.g.,][]{MacGregor1979, Owocki1984, Owocki1988}, and accretion instabilities in vicinity of a compact object.
For instance, for GX~301$-$2 the former scenario was considered by \cite{Mukherjee2004}.
On the other hand, recently \citet{ElMellah2020} suggested that the clumpiness of the stellar wind is likely not sufficient to explain such a large $N_{\rm H}^{\rm loc}$ variation.

Alternatively, if the $N_{\rm H}^{\rm loc}$ is mainly contributed by the material close to the neutron star (see below), the $N_{\rm H}^{\rm loc}$ variability could originate from some instabilities in accretion processes appearing in hydrodynamic simulations \citep{1991ApJ...371..684B,2014EPJWC..6402006M,ElMellah2020}.
The dynamical time-scale of the accretion is $R_{\rm acc}/v$ $\sim$6\,ks, which is well consistent with the observed timescales of both the flux and $N_{\rm H}^{\rm loc}$ variability \citep{Furst2011}. $R_{\rm acc} = \frac{2GM}{v^2} \sim 2.5\times10^9\,{\rm m}$ is the accretion radius, i.e., the impact parameter of streamlines gravitationally beamed by the neutron star.
Here $G$ is the gravitational constant, $M$ is the mass of the neutron star, and $v$ is the velocity of the flow before it is altered by the neutron star \citep{Edgar2004}.
The $v$ is a function of the terminal velocity ($v_{\infty}$) of the stellar wind and the velocity of the orbital movement.
In GX 301-2, the $v$ is $\sim$ 400\,km/s around the periastron \citep{Doroshenko2010}.

We note that the absorbing material is also expected to be responsible for reprocessing of X-ray emission, and in particular contribute to the formation of fluorescent iron lines.
Thanks to the good energy resolution of \textit{Insight}-HXMT/LE and its high cadence observational strategy, we are able to study the correlation of the $N_{\rm H}^{\rm loc}$ with and the equivalent width of iron $\rm k\alpha$ lines ($\rm EW_{\rm Fe\,\alpha}$), at different orbital phases.
We find that their correlation is well consistent with the theoretical prediction, if assuming that the iron line originates from X-ray reprocessing of hard X-rays with the accretion material surrounding the neutron star in approximately spherically symmetric geometry. This strongly suggests that same material is responsible both for absorption and X-ray reprocessing, which is consistent with results in supergiant X-ray binaries as reported by \citet{Gimenez2015}. The origin of the iron line (i.e., whether it forms close to the neutron star or on larger scales), however, is also still uncertain, and we discuss two possibilities here.

Based on the observed line width, \citet{Endo2002} proposed that it might be close to the Alfv\'en radius of the neutron star ($\sim10^8$\,m).
In this case, the approximate spherical configuration (resembling an "atmosphere") of accretion matter is expected \citep{Davies1979, Davies1981}, which is consistent with the observed $\rm EW_{\rm Fe\,\alpha}$-$N_{\rm H}^{\rm loc}$ relation revealed by our observations as described above.
On the other hand, \citet{Zheng2020} deduced somewhat longer distance (1.2$\pm$0.6$\times$ $10^{10}$\,m) to the iron line emitting region, based on the cross-correlation between light-curves of the iron line and the continuum.
If true, the $N_{H}$ variability might be caused by the large structure of the accretion stream.
However, there might be a problem that how to keep a spherical symmetry of matter at such a large scale. 

We note that, in either case, according to the $\rm EW_{\rm Fe\,\alpha}$-$N_{\rm H}^{\rm loc}$ relation, the variability of the $N_{\rm H}^{\rm loc}$ cannot be caused by a dense clump moving through the line of sight far away from the neutron star.
We note also that observed variations of the local absorption column are largely driven by short-term variations, and correlation of the iron line equivalent width and local absorption column also holds in this case as revealed both by \textit{Insight}-HXMT and \textit{XMM-Newton} observations during a flaring episode \citep[Fig. 12 in][]{Furst2011}.
This indicates that both the absorbing and re-processing material are located close to the neutron star and global spatial distribution of the accreting material does not play an important role in iron line and local absorption column properties, regardless on whether accretion proceeds from wind or accretion stream \citep[e.g.,][]{Leahy2008, Monkkonen2020}. Enhancement of the local absorption column close to the periastron passage coincident with the increase of X-ray flux could in this case be explained by increased local wind density around the neutron star.

It is interesting to note also that for spherically symmetric distribution of reprocessing material, a significantly reduced pulse fraction is predicted in the energy range around the Fe\,K$\alpha$ line complex. Indeed, in this case the X-ray reprocessing only produces non-pulsed radiation thereby reducing the total pulsed fraction. This inference is generally consistent with observations \citep[e.g., Fig. 5 in][]{Nabizadeh2019}. We note that a large size ($\gg$700-ls) of the line forming region suggested by \citet{Suchy2012} is therefore not required. On the other hand, \citet{Liu2018} reported pulsed iron line in a time interval of 7\,ks near the periastron, which disappeared afterwards. 
They explained that the intermittent pulsed iron line is associated with the anisotropic accretion, e.g., when going into or leaving the accretion stream. We note, however, that also in this case the asymmetry of reprocessing material may be local to the neutron star, i.e., be associated with an accretion wake or other features associated with interaction of the pulsar and the wind.

\section{Summary}
We studied the emission and absorption features of accreting X-ray pulsar GX 301-2 observed with \textit{Insight}-HXMT in 2017-2019.
We found prominent fluorescent lines of Fe K$\alpha$, Fe K$\beta$ and  Ni K$\alpha$, and the K-shell absorption of irons, in observations at its different orbital phases.
Our results show the capacity of \textit{Insight}-HXMT in the context of iron complex studies on several timescales associated with good energy resolution and the fast read-out that prevents pile-up effects.
We find that the Fe lines are orbital-dependent and their fluxes are correlated with those of the continuum.
We report on the first extensive study of the intensity ratio between Fe K$\alpha$ and K$\beta$ lines in GX 301-2. In particular, we find a linear coefficient is 0.12 $\pm$ 0.01 between them, which is in a good agreement with the theoretical prediction of irons with a low ionization state \citep{Palmeri2003}.

We find that in GX 301-2 the optical depth of the K-shell absorption of irons is correlated to  $\rm EW_{\rm Fe\,\alpha}$ with a linear coefficient of 2.5$\pm$0.4.
This coefficient is smaller than the result of \textit{Chandra}, which is $\sim$ 0.5 \citep{Torrejon2010}.
We note that the discrepancy might be caused by the influence of Fe K$\beta$ lines that cannot be resolved by \textit{Insight}-HXMT.

Compton shoulder previously reported based on the \textit{Chandra} data \citep{Watanabe2003, Torrejon2010, Tzanavaris2018} is significantly detected in several observations but find no clear relation with $N_{\rm H}^{\rm loc}$. We caution that, however, the energy resolution of \textit{Insight}-HXMT/LE is not sufficient to fully resolve shape of the iron line complex and observations with grating or micro-calorimeter instruments are required to verify our conclusion.

We find also strong variations of the observed absorption column $N_{\rm H}^{\rm loc}$ both on short and long timescales. In particular, the absorption is significantly enhanced around the periastron passage which is consistent with earlier studies \citep[e.g., ][]{Mukherjee2004,LaBarbera2005,Islam2014}.
For the first time, we studied also the relation between $N_{\rm H}^{\rm loc}$ and $\rm EW_{\rm Fe\,\alpha}$ measured at different orbital phases in GX 301-2. 
We find that it is in line with a theoretical estimation assuming that the accretion material is distributed spherically, regardless of specific accretion types, i.e., via stellar winds or a gas stream \citep{Leahy2008, Monkkonen2020}.	
We argue that together with rapid variations of both the $N_{\rm H}^{\rm loc}$ and iron line amplitude, this suggests that same material located close to the neutron star is likely responsible for both the absorption and X-ray reprocessing. We suggest, therefore, that rapid variability of the observed $N_{\rm H}^{\rm loc}$ is likely associated with inhomogenities and/or instabilities of the accretion flow around the neutron star. This conclusion is important in context of modelling of the observed orbital variation of X-ray flux and absorption column which shall thus be mostly considered as a tracer of local density around the pulsar rather than integral density of material along the line of sight.

\section*{Acknowledgements}
This work made use of the data from the \textit{Insight}-HXMT mission, a project funded by China National Space Administration (CNSA) and the Chinese Academy of Sciences (CAS). The \textit{Insight}-HXMT team gratefully acknowledges the support from the National Program on Key Research and Development Project (Grant No. 2016YFA0400800) from the Minister of Science and Technology of China (MOST) and the Strategic Priority Research Program of the Chinese Academy of Sciences (Grant No. XDB23040400). The authors thank supports from the National Natural Science Foundation of China under Grants No. 11503027, 11673023, 11733009, U1838201, U1838202, U1938103 and U2038101. JL, SV and NA thank the German Academic Exchange Service (DAAD, project57405000) for travel grants. VS, ST and VD acknowledge the support from the Russian Science Foundation grant 19-12-00423. VS thanks the Deutsche Forschungsgemeinschaft (DFG) grant WE 1312/51-1. MO acknowledges support from the Italian Space Agency under grant ASI-INAF 2017-14-H.0.





\section{Data Availability}
The data that support the findings of this study are available from \textit{Insight}-HMXT's data archive\footnote{\url{http://enghxmt.ihep.ac.cn/}} and the \textit{Swift}/BAT transient monitor\footnote{\url{https://swift.gsfc.nasa.gov/results/transients/}}.

\bibliographystyle{mnras}
\bibliography{mybib} 

\begin{thebibliography}{}
\makeatletter
\relax
\def\mn@urlcharsother{\let\do\@makeother \do\$\do\&\do\#\do\^\do\_\do\%\do\~}
\def\mn@doi{\begingroup\mn@urlcharsother \@ifnextchar [ {\mn@doi@}
  {\mn@doi@[]}}
\def\mn@doi@[#1]#2{\def\@tempa{#1}\ifx\@tempa\@empty \href
  {http://dx.doi.org/#2} {doi:#2}\else \href {http://dx.doi.org/#2} {#1}\fi
  \endgroup}
\def\mn@eprint#1#2{\mn@eprint@#1:#2::\@nil}
\def\mn@eprint@arXiv#1{\href {http://arxiv.org/abs/#1} {{\tt arXiv:#1}}}
\def\mn@eprint@dblp#1{\href {http://dblp.uni-trier.de/rec/bibtex/#1.xml}
  {dblp:#1}}
\def\mn@eprint@#1:#2:#3:#4\@nil{\def\@tempa {#1}\def\@tempb {#2}\def\@tempc
  {#3}\ifx \@tempc \@empty \let \@tempc \@tempb \let \@tempb \@tempa \fi \ifx
  \@tempb \@empty \def\@tempb {arXiv}\fi \@ifundefined
  {mn@eprint@\@tempb}{\@tempb:\@tempc}{\expandafter \expandafter \csname
  mn@eprint@\@tempb\endcsname \expandafter{\@tempc}}}

\bibitem[\protect\citeauthoryear{{Aftab}, {Paul}  \& {Kretschmar}}{{Aftab}
  et~al.}{2019}]{Aftab2019}
{Aftab} N.,  {Paul} B.,   {Kretschmar} P.,  2019, \mn@doi [\apjs]
  {10.3847/1538-4365/ab2a77}, \href
  {https://ui.adsabs.harvard.edu/abs/2019ApJS..243...29A} {243, 29}

\bibitem[\protect\citeauthoryear{{Arnaud}}{{Arnaud}}{1996}]{Arnaud1996}
{Arnaud} K.~A.,  1996, in {Jacoby} G.~H.,  {Barnes} J.,  eds, , Astronomical
  Data Analysis Software and Systems V.
Astronomical Society of the Pacific Conference Series, Vol. 101, p.~17

\bibitem[\protect\citeauthoryear{{Blondin}, {Stevens}  \& {Kallman}}{{Blondin}
  et~al.}{1991}]{1991ApJ...371..684B}
{Blondin} J.~M.,  {Stevens} I.~R.,   {Kallman} T.~R.,  1991, \mn@doi [\apj]
  {10.1086/169934}, \href
  {https://ui.adsabs.harvard.edu/abs/1991ApJ...371..684B} {371, 684}

\bibitem[\protect\citeauthoryear{{Castor}, {Abbott}  \& {Klein}}{{Castor}
  et~al.}{1975}]{Castor1975}
{Castor} J.~I.,  {Abbott} D.~C.,   {Klein} R.~I.,  1975, \mn@doi [\apj]
  {10.1086/153315}, \href
  {https://ui.adsabs.harvard.edu/abs/1975ApJ...195..157C} {195, 157}

\bibitem[\protect\citeauthoryear{{Chen} et~al.,}{{Chen}
  et~al.}{2019}]{Chen2019}
{Chen} Y.,  et~al., 2019, arXiv e-prints, \href
  {https://ui.adsabs.harvard.edu/abs/2019arXiv191008319C} {p. arXiv:1910.08319}

\bibitem[\protect\citeauthoryear{{Davies} \& {Pringle}}{{Davies} \&
  {Pringle}}{1981}]{Davies1981}
{Davies} R.~E.,  {Pringle} J.~E.,  1981, \mn@doi [\mnras]
  {10.1093/mnras/196.2.209}, \href
  {https://ui.adsabs.harvard.edu/abs/1981MNRAS.196..209D} {196, 209}

\bibitem[\protect\citeauthoryear{{Davies}, {Fabian}  \& {Pringle}}{{Davies}
  et~al.}{1979}]{Davies1979}
{Davies} R.~E.,  {Fabian} A.~C.,   {Pringle} J.~E.,  1979, \mn@doi [\mnras]
  {10.1093/mnras/186.4.779}, \href
  {https://ui.adsabs.harvard.edu/abs/1979MNRAS.186..779D} {186, 779}

\bibitem[\protect\citeauthoryear{{Doroshenko}, {Santangelo}, {Suleimanov},
  {Kreykenbohm}, {Staubert}, {Ferrigno}  \& {Klochkov}}{{Doroshenko}
  et~al.}{2010}]{Doroshenko2010}
{Doroshenko} V.,  {Santangelo} A.,  {Suleimanov} V.,  {Kreykenbohm} I.,
  {Staubert} R.,  {Ferrigno} C.,   {Klochkov} D.,  2010, \mn@doi [\aap]
  {10.1051/0004-6361/200912951}, \href
  {https://ui.adsabs.harvard.edu/abs/2010A&A...515A..10D} {515, A10}

\bibitem[\protect\citeauthoryear{{Edgar}}{{Edgar}}{2004}]{Edgar2004}
{Edgar} R.,  2004, \mn@doi [\nar] {10.1016/j.newar.2004.06.001}, \href
  {https://ui.adsabs.harvard.edu/abs/2004NewAR..48..843E} {48, 843}

\bibitem[\protect\citeauthoryear{{El Mellah}, {Grinberg}, {Sundqvist},
  {Driessen}  \& {Leutenegger}}{{El Mellah} et~al.}{2020}]{ElMellah2020}
{El Mellah} I.,  {Grinberg} V.,  {Sundqvist} J.~O.,  {Driessen} F.~A.,
  {Leutenegger} M.~A.,  2020, arXiv e-prints, \href
  {https://ui.adsabs.harvard.edu/abs/2020arXiv200616216E} {p. arXiv:2006.16216}

\bibitem[\protect\citeauthoryear{{Endo}, {Ishida}, {Masai}, {Kunieda}, {Inoue}
  \& {Nagase}}{{Endo} et~al.}{2002}]{Endo2002}
{Endo} T.,  {Ishida} M.,  {Masai} K.,  {Kunieda} H.,  {Inoue} H.,   {Nagase}
  F.,  2002, \mn@doi [\apj] {10.1086/341060}, \href
  {https://ui.adsabs.harvard.edu/abs/2002ApJ...574..879E} {574, 879}

\bibitem[\protect\citeauthoryear{{F{\"u}rst} et~al.,}{{F{\"u}rst}
  et~al.}{2011}]{Furst2011}
{F{\"u}rst} F.,  et~al., 2011, \mn@doi [\aap] {10.1051/0004-6361/201117665},
  \href {https://ui.adsabs.harvard.edu/abs/2011A&A...535A...9F} {535, A9}

\bibitem[\protect\citeauthoryear{{F{\"u}rst} et~al.,}{{F{\"u}rst}
  et~al.}{2018}]{Furst2018}
{F{\"u}rst} F.,  et~al., 2018, \mn@doi [\aap] {10.1051/0004-6361/201732132},
  \href {https://ui.adsabs.harvard.edu/abs/2018A&A...620A.153F} {620, A153}

\bibitem[\protect\citeauthoryear{{Gim{\'e}nez-Garc{\'\i}a}, {Torrej{\'o}n},
  {Eikmann}, {Mart{\'\i}nez-N{\'u}{\~n}ez}, {Oskinova}, {Rodes-Roca}  \&
  {Bernab{\'e}u}}{{Gim{\'e}nez-Garc{\'\i}a} et~al.}{2015}]{Gimenez2015}
{Gim{\'e}nez-Garc{\'\i}a} A.,  {Torrej{\'o}n} J.~M.,  {Eikmann} W.,
  {Mart{\'\i}nez-N{\'u}{\~n}ez} S.,  {Oskinova} L.~M.,  {Rodes-Roca} J.~J.,
  {Bernab{\'e}u} G.,  2015, \mn@doi [\aap] {10.1051/0004-6361/201425004}, \href
  {https://ui.adsabs.harvard.edu/abs/2015A&A...576A.108G} {576, A108}

\bibitem[\protect\citeauthoryear{{HI4PI Collaboration} et~al.,}{{HI4PI
  Collaboration} et~al.}{2016}]{HI4PI2016}
{HI4PI Collaboration} et~al., 2016, \mn@doi [\aap]
  {10.1051/0004-6361/201629178}, \href
  {https://ui.adsabs.harvard.edu/abs/2016A&A...594A.116H} {594, A116}

\bibitem[\protect\citeauthoryear{{Haberl}}{{Haberl}}{1991}]{Haberl1991}
{Haberl} F.,  1991, \mn@doi [\apj] {10.1086/170273}, \href
  {https://ui.adsabs.harvard.edu/abs/1991ApJ...376..245H} {376, 245}

\bibitem[\protect\citeauthoryear{{Inoue}}{{Inoue}}{1985}]{Inoue1985}
{Inoue} H.,  1985, \mn@doi [\ssr] {10.1007/BF00212905}, \href
  {https://ui.adsabs.harvard.edu/abs/1985SSRv...40..317I} {40, 317}

\bibitem[\protect\citeauthoryear{{Islam} \& {Paul}}{{Islam} \&
  {Paul}}{2014}]{Islam2014}
{Islam} N.,  {Paul} B.,  2014, \mn@doi [\mnras] {10.1093/mnras/stu756}, \href
  {https://ui.adsabs.harvard.edu/abs/2014MNRAS.441.2539I} {441, 2539}

\bibitem[\protect\citeauthoryear{{Kallman}, {Palmeri}, {Bautista}, {Mendoza}
  \& {Krolik}}{{Kallman} et~al.}{2004}]{Kallman2004}
{Kallman} T.~R.,  {Palmeri} P.,  {Bautista} M.~A.,  {Mendoza} C.,   {Krolik}
  J.~H.,  2004, \mn@doi [\apjs] {10.1086/424039}, \href
  {https://ui.adsabs.harvard.edu/abs/2004ApJS..155..675K} {155, 675}

\bibitem[\protect\citeauthoryear{{Kaper}, {Lamers}, {Ruymaekers}, {van den
  Heuvel}  \& {Zuiderwijk}}{{Kaper} et~al.}{1995}]{Kaper1995}
{Kaper} L.,  {Lamers} H.~J.~G.~L.~M.,  {Ruymaekers} E.,  {van den Heuvel}
  E.~P.~J.,   {Zuiderwijk} E.~J.,  1995, \aap, \href
  {https://ui.adsabs.harvard.edu/abs/1995A&A...300..446K} {300, 446}

\bibitem[\protect\citeauthoryear{{Kaper}, {van der Meer}  \& {Najarro}}{{Kaper}
  et~al.}{2006}]{Kaper2006}
{Kaper} L.,  {van der Meer} A.,   {Najarro} F.,  2006, \mn@doi [\aap]
  {10.1051/0004-6361:20065393}, \href
  {https://ui.adsabs.harvard.edu/abs/2006A&A...457..595K} {457, 595}

\bibitem[\protect\citeauthoryear{{Koh} et~al.,}{{Koh} et~al.}{1997}]{Koh1997}
{Koh} D.~T.,  et~al., 1997, \mn@doi [\apj] {10.1086/303929}, \href
  {https://ui.adsabs.harvard.edu/abs/1997ApJ...479..933K} {479, 933}

\bibitem[\protect\citeauthoryear{{Kreykenbohm}, {Wilms}, {Coburn}, {Kuster},
  {Rothschild}, {Heindl}, {Kretschmar}  \& {Staubert}}{{Kreykenbohm}
  et~al.}{2004}]{Kreykenbohm2004}
{Kreykenbohm} I.,  {Wilms} J.,  {Coburn} W.,  {Kuster} M.,  {Rothschild} R.~E.,
   {Heindl} W.~A.,  {Kretschmar} P.,   {Staubert} R.,  2004, \mn@doi [\aap]
  {10.1051/0004-6361:20035836}, \href
  {https://ui.adsabs.harvard.edu/abs/2004A&A...427..975K} {427, 975}

\bibitem[\protect\citeauthoryear{{La Barbera}, {Segreto}, {Santangelo},
  {Kreykenbohm}  \& {Orlandini}}{{La Barbera} et~al.}{2005}]{LaBarbera2005}
{La Barbera} A.,  {Segreto} A.,  {Santangelo} A.,  {Kreykenbohm} I.,
  {Orlandini} M.,  2005, \mn@doi [\aap] {10.1051/0004-6361:20041509}, \href
  {https://ui.adsabs.harvard.edu/abs/2005A&A...438..617L} {438, 617}

\bibitem[\protect\citeauthoryear{{Leahy}}{{Leahy}}{1991}]{Leahy1991}
{Leahy} D.~A.,  1991, \mn@doi [\mnras] {10.1093/mnras/250.2.310}, \href
  {https://ui.adsabs.harvard.edu/abs/1991MNRAS.250..310L} {250, 310}

\bibitem[\protect\citeauthoryear{{Leahy} \& {Kostka}}{{Leahy} \&
  {Kostka}}{2008}]{Leahy2008}
{Leahy} D.~A.,  {Kostka} M.,  2008, \mn@doi [\mnras]
  {10.1111/j.1365-2966.2007.12754.x}, \href
  {https://ui.adsabs.harvard.edu/abs/2008MNRAS.384..747L} {384, 747}

\bibitem[\protect\citeauthoryear{{Liu}, {Soria}, {Qiao}  \& {Liu}}{{Liu}
  et~al.}{2018}]{Liu2018}
{Liu} J.,  {Soria} R.,  {Qiao} E.,   {Liu} J.,  2018, \mn@doi [\mnras]
  {10.1093/mnras/sty2180}, \href
  {https://ui.adsabs.harvard.edu/abs/2018MNRAS.480.4746L} {480, 4746}

\bibitem[\protect\citeauthoryear{{MacGregor}, {Hartmann}  \&
  {Raymond}}{{MacGregor} et~al.}{1979}]{MacGregor1979}
{MacGregor} K.~B.,  {Hartmann} L.,   {Raymond} J.~C.,  1979, \mn@doi [\apj]
  {10.1086/157213}, \href
  {https://ui.adsabs.harvard.edu/abs/1979ApJ...231..514M} {231, 514}

\bibitem[\protect\citeauthoryear{{Manousakis}, {Walter}  \&
  {Blondin}}{{Manousakis} et~al.}{2014}]{2014EPJWC..6402006M}
{Manousakis} A.,  {Walter} R.,   {Blondin} J.,  2014, in European Physical
  Journal Web of Conferences. p. 02006 (\mn@eprint {arXiv} {1310.8205}),
  \mn@doi{10.1051/epjconf/20136402006}

\bibitem[\protect\citeauthoryear{{Matt}}{{Matt}}{2002}]{Matt2002}
{Matt} G.,  2002, \mn@doi [\mnras] {10.1046/j.1365-8711.2002.05890.x}, \href
  {https://ui.adsabs.harvard.edu/abs/2002MNRAS.337..147M} {337, 147}

\bibitem[\protect\citeauthoryear{{M{\"o}nkk{\"o}nen}, {Doroshenko},
  {Tsygankov}, {Nabizadeh}, {Abolmasov}  \& {Poutanen}}{{M{\"o}nkk{\"o}nen}
  et~al.}{2020}]{Monkkonen2020}
{M{\"o}nkk{\"o}nen} J.,  {Doroshenko} V.,  {Tsygankov} S.~S.,  {Nabizadeh} A.,
  {Abolmasov} P.,   {Poutanen} J.,  2020, \mn@doi [\mnras]
  {10.1093/mnras/staa906}, \href
  {https://ui.adsabs.harvard.edu/abs/2020MNRAS.494.2178M} {494, 2178}

\bibitem[\protect\citeauthoryear{{Mukherjee} \& {Paul}}{{Mukherjee} \&
  {Paul}}{2004}]{Mukherjee2004}
{Mukherjee} U.,  {Paul} B.,  2004, \mn@doi [\aap] {10.1051/0004-6361:20034407},
  \href {https://ui.adsabs.harvard.edu/abs/2004A&A...427..567M} {427, 567}

\bibitem[\protect\citeauthoryear{{Nabizadeh}, {M{\"o}nkk{\"o}nen}, {Tsygankov},
  {Doroshenko}, {Molkov}  \& {Poutanen}}{{Nabizadeh}
  et~al.}{2019}]{Nabizadeh2019}
{Nabizadeh} A.,  {M{\"o}nkk{\"o}nen} J.,  {Tsygankov} S.~S.,  {Doroshenko} V.,
  {Molkov} S.~V.,   {Poutanen} J.,  2019, \mn@doi [\aap]
  {10.1051/0004-6361/201936045}, \href
  {https://ui.adsabs.harvard.edu/abs/2019A&A...629A.101N} {629, A101}

\bibitem[\protect\citeauthoryear{{Owocki} \& {Rybicki}}{{Owocki} \&
  {Rybicki}}{1984}]{Owocki1984}
{Owocki} S.~P.,  {Rybicki} G.~B.,  1984, \mn@doi [\apj] {10.1086/162412}, \href
  {https://ui.adsabs.harvard.edu/abs/1984ApJ...284..337O} {284, 337}

\bibitem[\protect\citeauthoryear{{Owocki}, {Castor}  \& {Rybicki}}{{Owocki}
  et~al.}{1988}]{Owocki1988}
{Owocki} S.~P.,  {Castor} J.~I.,   {Rybicki} G.~B.,  1988, \mn@doi [\apj]
  {10.1086/166977}, \href
  {https://ui.adsabs.harvard.edu/abs/1988ApJ...335..914O} {335, 914}

\bibitem[\protect\citeauthoryear{{Palmeri}, {Mendoza}, {Kallman}, {Bautista}
  \& {Mel{\'e}ndez}}{{Palmeri} et~al.}{2003}]{Palmeri2003}
{Palmeri} P.,  {Mendoza} C.,  {Kallman} T.~R.,  {Bautista} M.~A.,
  {Mel{\'e}ndez} M.,  2003, \mn@doi [\aap] {10.1051/0004-6361:20031262}, \href
  {https://ui.adsabs.harvard.edu/abs/2003A&A...410..359P} {410, 359}

\bibitem[\protect\citeauthoryear{{Pravdo}, {Day}, {Angelini}, {Harmon},
  {Yoshida}  \& {Saraswat}}{{Pravdo} et~al.}{1995}]{Pravdo1995}
{Pravdo} S.~H.,  {Day} C. S.~R.,  {Angelini} L.,  {Harmon} B.~A.,  {Yoshida}
  A.,   {Saraswat} P.,  1995, \mn@doi [\apj] {10.1086/176540}, \href
  {https://ui.adsabs.harvard.edu/abs/1995ApJ...454..872P} {454, 872}

\bibitem[\protect\citeauthoryear{{Rothschild} \& {Soong}}{{Rothschild} \&
  {Soong}}{1987}]{Rothschild1987}
{Rothschild} R.~E.,  {Soong} Y.,  1987, \mn@doi [\apj] {10.1086/165121}, \href
  {https://ui.adsabs.harvard.edu/abs/1987ApJ...315..154R} {315, 154}

\bibitem[\protect\citeauthoryear{{Sato}, {Nagase}, {Kawai}, {Kelley},
  {Rappaport}  \& {White}}{{Sato} et~al.}{1986}]{Sato1986}
{Sato} N.,  {Nagase} F.,  {Kawai} N.,  {Kelley} R.~L.,  {Rappaport} S.,
  {White} N.~E.,  1986, \mn@doi [\apj] {10.1086/164157}, \href
  {https://ui.adsabs.harvard.edu/abs/1986ApJ...304..241S} {304, 241}

\bibitem[\protect\citeauthoryear{{Staubert} et~al.,}{{Staubert}
  et~al.}{2019}]{Staubert2019}
{Staubert} R.,  et~al., 2019, \mn@doi [\aap] {10.1051/0004-6361/201834479},
  \href {https://ui.adsabs.harvard.edu/abs/2019A&A...622A..61S} {622, A61}

\bibitem[\protect\citeauthoryear{{Stevens}}{{Stevens}}{1988}]{Stevens1988}
{Stevens} I.~R.,  1988, \mn@doi [\mnras] {10.1093/mnras/235.2.523}, \href
  {https://ui.adsabs.harvard.edu/abs/1988MNRAS.235..523S} {235, 523}

\bibitem[\protect\citeauthoryear{{Suchy}, {F{\"u}rst}, {Pottschmidt},
  {Caballero}, {Kreykenbohm}, {Wilms}, {Markowitz}  \& {Rothschild}}{{Suchy}
  et~al.}{2012}]{Suchy2012}
{Suchy} S.,  {F{\"u}rst} F.,  {Pottschmidt} K.,  {Caballero} I.,  {Kreykenbohm}
  I.,  {Wilms} J.,  {Markowitz} A.,   {Rothschild} R.~E.,  2012, \mn@doi [\apj]
  {10.1088/0004-637X/745/2/124}, \href
  {https://ui.adsabs.harvard.edu/abs/2012ApJ...745..124S} {745, 124}

\bibitem[\protect\citeauthoryear{{Torrej{\'o}n}, {Schulz}, {Nowak}  \&
  {Kallman}}{{Torrej{\'o}n} et~al.}{2010}]{Torrejon2010}
{Torrej{\'o}n} J.~M.,  {Schulz} N.~S.,  {Nowak} M.~A.,   {Kallman} T.~R.,
  2010, \mn@doi [\apj] {10.1088/0004-637X/715/2/947}, \href
  {https://ui.adsabs.harvard.edu/abs/2010ApJ...715..947T} {715, 947}

\bibitem[\protect\citeauthoryear{{Tzanavaris} \& {Yaqoob}}{{Tzanavaris} \&
  {Yaqoob}}{2018}]{Tzanavaris2018}
{Tzanavaris} P.,  {Yaqoob} T.,  2018, \mn@doi [\apj]
  {10.3847/1538-4357/aaaab6}, \href
  {https://ui.adsabs.harvard.edu/abs/2018ApJ...855...25T} {855, 25}

\bibitem[\protect\citeauthoryear{{Vidal}}{{Vidal}}{1973}]{Vidal1973}
{Vidal} N.~V.,  1973, \mn@doi [\apjl] {10.1086/181362}, \href
  {https://ui.adsabs.harvard.edu/abs/1973ApJ...186L..81V} {186, L81}

\bibitem[\protect\citeauthoryear{{Waisberg} et~al.,}{{Waisberg}
  et~al.}{2017}]{2017ApJ...844...72W}
{Waisberg} I.,  et~al., 2017, \mn@doi [\apj] {10.3847/1538-4357/aa79f1}, \href
  {https://ui.adsabs.harvard.edu/abs/2017ApJ...844...72W} {844, 72}

\bibitem[\protect\citeauthoryear{{Watanabe} et~al.,}{{Watanabe}
  et~al.}{2003}]{Watanabe2003}
{Watanabe} S.,  et~al., 2003, \mn@doi [\apjl] {10.1086/379735}, \href
  {https://ui.adsabs.harvard.edu/abs/2003ApJ...597L..37W} {597, L37}

\bibitem[\protect\citeauthoryear{{Wilms}, {Allen}  \& {McCray}}{{Wilms}
  et~al.}{2000}]{Wilms2000}
{Wilms} J.,  {Allen} A.,   {McCray} R.,  2000, \mn@doi [\apj] {10.1086/317016},
  \href {https://ui.adsabs.harvard.edu/abs/2000ApJ...542..914W} {542, 914}

\bibitem[\protect\citeauthoryear{{Zhang} et~al.,}{{Zhang}
  et~al.}{2014}]{Zhang2014}
{Zhang} S.,  et~al., 2014, \mn@doi [\procspie] {10.1117/12.2054131}, \href
  {https://ui.adsabs.harvard.edu/abs/2014SPIE.9144E..55Z} {9144, 914455}

\bibitem[\protect\citeauthoryear{{Zhang} et~al.,}{{Zhang}
  et~al.}{2019}]{Zhang2019}
{Zhang} S.,  et~al., 2019, arXiv e-prints, \href
  {https://ui.adsabs.harvard.edu/abs/2019arXiv191009613Z} {p. arXiv:1910.09613}

\bibitem[\protect\citeauthoryear{{Zheng}, {Liu}  \& {Gou}}{{Zheng}
  et~al.}{2020}]{Zheng2020}
{Zheng} X.,  {Liu} J.,   {Gou} L.,  2020, \mn@doi [\mnras]
  {10.1093/mnras/stz3327}, \href
  {https://ui.adsabs.harvard.edu/abs/2020MNRAS.491.4802Z} {491, 4802}

\makeatother
\end{thebibliography}

\begin{table*}
\small
\caption{The columns denote observational IDs, the observation date, the exposure, the orbital phase assuming the ephemeris adopted from \citet{Koh1997}, and best-fitting parameters of the continuum that is described as an absorbed power-law model. }
\begin{tabular}{ccccccc}
\hline
ObsID & Time & Exposure & $\rm Phase_{orb}$ & $\Gamma$ & $\rm N_{H}^{\rm loc}$ & Unabsorbed\ $\rm Flux_ {5.5-8.5\,keV}$ \\ 
	& (MJD) & (s) &  &  &  $(10^{22}$ ${\rm atoms\,cm^{-2}})$  &  ($\rm 10^{-9}\,erg\,cm^{-2}\,s^{-1}$) \\
\hline
P010130900101 & 57968.33 & 1886 & 0.87 & $0.95_{-0.05}^{+0.06}$ & $95.64_{-4.84}^{+4.04}$   & $6.39_{-0.62}^{+0.57}$    \\ 
P010130900102 & 57968.46 & 2027 & 0.87 & $1.22_{-0.09}^{+0.13}$ & $93.57_{-7.61}^{+4.25}$   & $3.81_{-0.55}^{+0.35}$    \\ 
P010130900103 & 57968.60 & 1792 & 0.87 & $1.31_{-0.06}^{+0.07}$ & $125.18_{-5.83}^{+4.62}$  & $9.18_{-1.05}^{+0.95}$     \\ 
P010130900104 & 57968.73 & 1506 & 0.88 & $2.53_{-0.12}^{+0.09}$ & $211.48_{-6.05}^{+4.74}$  & $11.65_{-1.69}^{+1.19}$    \\ 
P010130900107 & 57969.15 & 2538 & 0.89 & $1.29_{-0.08}^{+0.08}$ & $114.35_{-5.28}^{+6.09}$  & $4.30_{-0.45}^{+0.59}$     \\ 
P010130900401 & 57969.58 & 1827 & 0.90 & $1.89_{-0.10}^{+0.07}$ & $128.83_{-5.87}^{+8.45}$  & $2.53_{-0.31}^{+0.45}$     \\ 
P010130900402 & 57969.73 & 1264 & 0.90 & $1.15_{-0.06}^{+0.03}$ & $66.68_{-3.66}^{+2.02}$   & $4.16_{-0.34}^{+0.15}$    \\ 
P010130900502 & 58121.32 & 1088 & 0.55 & $2.10_{-0.04}^{+0.06}$ & $18.84_{-1.99}^{+3.06}$   & $0.41_{-0.01}^{+0.02}$    \\ 
P010130900701 & 58137.54 & 1036 & 0.94 & $1.73_{-0.03}^{+0.02}$ & $80.65_{-1.27}^{+0.31}$   & $0.95_{-0.02}^{+0.02}$    \\ 
P010130900808 & 58148.71 & 2513 & 0.21 & $1.58_{-0.01}^{+0.01}$ & $57.67_{-0.39}^{+0.52}$   & $0.63_{-0.01}^{+0.01}$    \\ 
P010130900901 & 58163.01 & 2914 & 0.56 & $2.45_{-0.03}^{+0.02}$ & $50.16_{-3.82}^{+3.40}$   & $0.46_{-0.04}^{+0.05}$    \\ 
P010130901501 & 58218.03 & 7713 & 0.88 & $1.33_{-0.07}^{+0.06}$ & $150.24_{-6.46}^{+4.73}$  & $7.74_{-0.96}^{+0.81}$     \\ 
P010130901601 & 58258.34 & 1645 & 0.86 & $1.13_{-0.06}^{+0.08}$ & $113.93_{-6.07}^{+5.38}$  & $7.11_{-0.87}^{+0.87}$     \\ 
P010130901602 & 58258.50 & 1188 & 0.86 & $1.84_{-0.14}^{+0.09}$ & $201.33_{-5.54}^{+5.39}$  & $13.02_{-1.67}^{+2.01}$    \\ 
P010130901701 & 58259.83 & 1012 & 0.89 & $2.51_{-0.14}^{+0.19}$ & $167.21_{-16.04}^{+14.86}$& $3.95_{-1.04}^{+1.38}$     \\
P010130901702 & 58259.96 & 1182 & 0.89 & $1.99_{-0.17}^{+0.15}$ & $170.67_{-11.10}^{+8.03}$ & $4.25_{-1.10}^{+1.08}$     \\ 
P010130901802 & 58327.04 & 1206 & 0.51 & $1.67_{-0.03}^{+0.03}$ & $30.82_{-2.32}^{+2.00}$   & $0.40_{-0.03}^{+0.03}$     \\ 
P010130901901 & 58494.92 & 2685 & 0.56 & $1.42_{-0.02}^{+0.03}$ & $35.83_{-1.49}^{+2.27}$   & $1.98_{-0.07}^{+0.11}$     \\ 
P010130902001 & 58495.75 & 3432 & 0.58 & $1.48_{-0.01}^{+0.02}$ & $47.04_{-0.42}^{+1.45}$   & $2.38_{-0.05}^{+0.06}$     \\ 
P010130902101 & 58545.36 & 2092 & 0.77 & $2.00_{-0.10}^{+0.08}$ & $41.14_{-3.71}^{+4.12}$   & $0.69_{-0.06}^{+0.05}$     \\ 
P010130902102 & 58545.49 & 2214 & 0.78 & $1.54_{-0.01}^{+0.01}$ & $40.69_{-0.50}^{+0.30}$   & $1.06_{-0.01}^{+0.01}$     \\ 
P010130902103 & 58545.63 & 2495 & 0.78 & $1.46_{-0.03}^{+0.04}$ & $54.69_{-2.74}^{+2.97}$   & $2.34_{-0.14}^{+0.14}$     \\ 
P010130902104 & 58545.79 & 2894 & 0.78 & $1.28_{-0.03}^{+0.02}$ & $51.10_{-4.07}^{+2.67}$   & $2.61_{-0.21}^{+0.13}$     \\ 
\hline
\end{tabular}
\label{tab:ObsID}
\end{table*}

\newpage
\begin{sidewaystable}
\vspace{-20cm}
\tiny
\centering
\caption{The parameters of emission and absorption features, i.e., the $\rm Fe\  K\alpha$, $\rm Fe\ K\beta$, $\rm Ni\ K\alpha$, CS and the optical depth of the iron K shell, where "-" represents features that can not be identified at 3\,$\sigma$ confidence level.}
\begin{tabular}{cccccccccccc}
\hline
ObsID  & $\tau_{\rm K\,edge}$ & $\rm E_{Fe\,K\alpha}$ & $\rm EW_{\rm  Fe\,K\alpha}$ & $\rm I_{Fe\,K\alpha}$& $\rm E_{\rm Fe\,K\beta}$ & $\rm EW_{\rm Fe\,K\beta}$ & $\rm I_{Fe\,K\beta}$& $\rm E_{\rm Ni\,K\alpha}$ & $\rm EW_{\rm Ni\,K\alpha}$ &$\rm I_{Ni\,K\alpha}$& $\rm EW_{\rm CS}$\\
& (MJD)& (keV) & (eV) & ($\rm 10^{-3}\,photons/cm^2/s$) & (keV) & (eV) & ($\rm 10^{-3}\,photons/cm^2/s$) & (keV) & (eV) & ($\rm 10^{-3}\,photons/cm^2/s$) & (eV) \\
\hline
P010130900101 & $0.89_{-0.03}^{+0.03}$ & $6.406_{-0.002}^{+0.002}$ & $400.86_{-19.52}^{+23.18}$    & $14.51_{-0.47}^{+0.50}$ & $7.05_{-0.02}^{+0.02}$ & $64.64_{-4.47}^{+4.74}$    & $1.87_{-0.13}^{+0.14}$ & $7.48_{-0.02}^{+0.02}$ & $42.13_{-11.37}^{+15.68}$ & $0.82_{-0.22}^{+0.30}$ & $58.91_{-9.20}^{+8.94}$    \\
P010130900102 & $0.87_{-0.06}^{+0.05}$ & $6.398_{-0.003}^{+0.001}$ & $433.10_{-44.96}^{+25.67}$    & $8.92_{-0.53}^{+0.30}$  & $6.99_{-0.05}^{+0.02}$ & $66.68_{-11.47}^{+10.98}$  & $1.24_{-0.16}^{+0.18}$ & -                      & -                         & -                      & -      \\
P010130900103 & $1.18_{-0.06}^{+0.04}$ & $6.404_{-0.004}^{+0.003}$ & $467.61_{-30.70}^{+50.95}$    & $13.46_{-0.58}^{+0.73}$ & $7.00_{-0.03}^{+0.03}$ & $58.84_{-7.12}^{+8.63}$    & $1.46_{-0.16}^{+0.21}$ & $7.50_{-0.02}^{+0.03}$ & $55.73_{-15.75}^{+19.02}$ & $0.76_{-0.21}^{+0.26}$ & $40.08_{-21.98}^{+10.53}$  \\
P010130900104 & -                      & $6.396_{-0.004}^{+0.006}$ & $1064.29_{-240.46}^{+194.36}$ & $8.33_{-0.85}^{+0.54}$  & $7.05_{-0.01}^{+0.01}$ & $215.14_{-26.55}^{+50.99}$ & $1.30_{-0.13}^{+0.23}$ & -                      & -                         & -                      & $42.97_{-28.41}^{+60.68}$  \\
P010130900107 & $0.98_{-0.05}^{+0.05}$ & $6.400_{-0.003}^{+0.007}$ & $442.36_{-53.19}^{+29.14}$    & $6.97_{-0.51}^{+0.24}$  & $7.02_{-0.04}^{+0.03}$ & $37.94_{-14.54}^{+13.60}$  & $0.53_{-0.19}^{+0.17}$ & $7.55_{-0.04}^{+0.04}$ & $79.51_{-22.52}^{+22.22}$ & $0.66_{-0.20}^{+0.18}$ & -      \\
P010130900401 & -                      & $6.402_{-0.007}^{+0.014}$ & $495.75_{-120.75}^{+65.23}$   & $3.74_{-0.56}^{+0.28}$  & $6.99_{-0.04}^{+0.03}$ & $86.14_{-16.97}^{+33.57}$  & $0.55_{-0.10}^{+0.19}$ & -                      & -                         & -                      & -      \\
P010130900402 & $0.68_{-0.06}^{+0.06}$ & $6.401_{-0.008}^{+0.007}$ & $235.38_{-1.73}^{+1.94}$      & $9.08_{-0.04}^{+0.04}$  & $6.95_{-0.02}^{+0.03}$ & $32.81_{-5.45}^{+10.99}$   & $1.13_{-0.19}^{+0.35}$ & -                      & -                         & -                      & -      \\
P010130900502 & -                      & $6.430_{-0.035}^{+0.039}$ & $123.02_{-4.63}^{+6.78}$      & $1.23_{-0.04}^{+0.04}$  & -                      & -                          & -                      & -                      & -                         & -                      & -      \\
P010130900701 & $0.70_{-0.01}^{+0.01}$ & $6.404_{-0.022}^{+0.020}$ & $295.17_{-13.56}^{+3.30}$     & $1.93_{-0.02}^{+0.04}$  & -                      & -                          & -                      & -                      & -                         & -                      & -      \\
P010130900808 & -                      & -                         & -                             & -                       & -                      & -                          & -                      & -                      & -                         & -                      & -      \\
P010130900901 & $0.44_{-0.01}^{+0.01}$ & -                         & -                             & -                       & -                      & -                          & -                      & -                      & -                         & -                      & -      \\
P010130901501 & $1.40_{-0.01}^{+0.01}$ & $6.401_{-0.002}^{+0.005}$ & $505.54_{-57.16}^{+26.39}$    & $8.65_{-0.56}^{+0.20}$  & $7.00_{-0.02}^{+0.04}$ & $64.79_{-4.79}^{+6.09}$    & $0.85_{-0.06}^{+0.07}$ & -                      & -                         & -                      & $95.17_{-17.78}^{+22.09}$  \\
P010130901601 & $0.92_{-0.01}^{+0.01}$ & $6.419_{-0.004}^{+0.008}$ & $253.64_{-24.17}^{+14.80}$    & $8.14_{-0.53}^{+0.35}$  & $6.95_{-0.03}^{+0.04}$ & $56.02_{-9.78}^{+9.05}$    & $1.38_{-0.22}^{+0.21}$ & $7.57_{-0.03}^{+0.03}$ & $48.06_{-18.41}^{+21.35}$ & $0.68_{-0.25}^{+0.30}$ & $132.77_{-14.12}^{+16.62}$ \\
P010130901602 & -                      & $6.410_{-0.004}^{+0.008}$ & $809.40_{-170.49}^{+145.68}$  & $9.52_{-0.95}^{+0.78}$  & $7.07_{-0.03}^{+0.03}$ & $174.01_{-28.41}^{+30.09}$ & $1.40_{-0.23}^{+0.21}$ & -                      & -                         & -                      & $84.41_{-36.60}^{+65.98}$  \\
P010130901701 & -                      & $6.417_{-0.005}^{+0.003}$ & $781.20_{-111.72}^{+210.39}$  & $6.96_{-0.47}^{+0.73}$  & $7.05_{-0.03}^{+0.03}$ & $249.79_{-30.64}^{+31.16}$ & $1.17_{-0.16}^{+0.13}$ & -                      & -                         & -                      & $143.52_{-46.35}^{+33.87}$ \\
P010130901702 & -                      & $6.395_{-0.008}^{+0.008}$ & $886.00_{-270.04}^{+326.83}$  & $6.42_{-0.97}^{+0.88}$  & $7.00_{-0.01}^{+0.02}$ & $211.27_{-67.37}^{+91.89}$ & $1.04_{-0.27}^{+0.33}$ & -                      & -                         & -                      & $91.75_{-68.99}^{+110.96}$ \\
P010130901802 & -                      & -                         & -                             & -                       & -                      & -                          & -                      & -                      & -                         & -                      & -      \\
P010130901901 & $0.35_{-0.04}^{+0.06}$ & $6.419_{-0.003}^{+0.019}$ & $104.02_{-1.08}^{+1.00}$      & $3.49_{-0.03}^{+0.06}$  & -                      & -                          & -                      & -                      & -                         & -                      & -      \\
P010130902001 & $0.33_{-0.06}^{+0.04}$ & $6.396_{-0.015}^{+0.013}$ & $108.07_{-14.49}^{+3.67}$     & $3.41_{-0.37}^{+0.09}$  & $7.03_{-0.01}^{+0.01}$ & $16.49_{-1.29}^{+0.38}$    & $0.45_{-0.03}^{+0.01}$ & -                      & -                         & -                      & -      \\
P010130902101 & -                      & $6.452_{-0.029}^{+0.031}$ & $125.40_{-4.31}^{+4.45}$      & $1.46_{-0.05}^{+0.04}$  & -                      & -                          & -                      & -                      & -                         & -                      & -      \\
P010130902102 & $0.47_{-0.01}^{+0.02}$ & $6.415_{-0.009}^{+0.011}$ & $203.40_{-1.20}^{+1.47}$      & $3.23_{-0.00}^{+0.00}$  & -                      & -                          & -                      & -                      & -                         & -                      & -      \\
P010130902103 & $0.45_{-0.05}^{+0.06}$ & $6.428_{-0.020}^{+0.026}$ & $145.22_{-34.21}^{+31.36}$    & $4.07_{-0.80}^{+0.68}$  & -                      & -                          & -                      & -                      & -                         & -                      & -      \\
P010130902104 & $0.41_{-0.02}^{+0.02}$ & $6.431_{-0.013}^{+0.012}$ & $146.87_{-17.03}^{+20.20}$    & $5.29_{-0.52}^{+0.57}$  & $7.12_{-0.04}^{+0.04}$ & $20.69_{-6.30}^{+5.36}$    & $0.56_{-0.18}^{+0.14}$ & -                      & -                         & -                      & $77.85_{-16.62}^{+13.12}$  \\
\hline
\end{tabular}
\label{tab:par1}
\end{sidewaystable}







\bsp	
\label{lastpage}
\end{document}